# On the solution of the Graph Isomorphism Problem

# Part I

Leonid I. Malinin
Natalia L. Malinina

**Abstract**

The presented matirial is devoted to the equivalent conversion from the vertex graphs to the edge graphs. We suggest that the proved theorems solve the problem of the isomorphism of graphs, the problem of the graph's enumeration with the help of the effective algorithms without their preliminary plotting, etc. The examining of the transformation of the vertex graphs into the edge graphs illustrates the reasons of the appearance of NP-completeness from the point of view of the graph theory. We suggest that it also illustrates the synchronous possibility and impossibility of the struggle with NP-completeness.

## I. An equivalent conversion between the graphs

### 1 Introduction. A duality of both the vertex and the edge graphs

Both the edge and the vertex graphs at the fixed conditions turn out to be dual. In graph theory the duality appears to be between the vertexes of one graph and the edges of the other graph and, at the same time, between the edges of one graph and the edges of the other graph [1, 2]. We are interested in the duality between the edges of one graph and the vertexes of the other graph, and we want such duality to be the reversible one. Let us investigate if there is a duality between the edges of one directed graph and the vertexes of the other one.

Such type of the duality for the undirected graphs was examined in full by other authors in [1, 2, 3, 4] and so forth. On the whole such duality comes to the fact that any undirected $H$ graph has the $G$ graph, which is dual to it:

- Every edge of the $H$ graph corresponds to every vertex of the $G$ graph.

- In the $G$ graph only those vertexes are joined by the edges that correspond to the edges in the $H$ graph, which are joined by means of the vertexes.

The opposite, broadly speaking, does not exist [2].

A $G$ graph is denoted as the adjacency graph of the $H$ graph's edges [2]. A $G$ graph is also denoted as either the derived graph [2] or the edge graph [1, 4]. An elegant criterion of the $G$ graph being the adjacency graph of the $H$ graph's edges was suggested by Krausz in [4]. It was displayed that the $G$ graph appeared to be the adjacency graph of the $H$ graph's edges if and only if such partition of the $G$ graph's set into the complete subgraphs exists in such a way that no one of the $G$ graph's vertexes are situated in more than two such subgraphs [2, 5]. This criterion allows answering whether the $H$ graph exists, if the $G$ graph is specified. But his criterion proves to be useful only for the undirected graphs.

But we are interested in the duality between the directed graphs and such duality was rarely investigated. Let's denote the directed $H$ graph as edge graph, and the directed $G$ graph as vertex graph.

A matrix of either the direct paths or the binary relations contains only directed connections (including the pairs of mutually opposite directed pairs). So, all the operations, which are accomplished on the matrix of the direct paths, are the operations, which are accomplished on the directed graph.

Let us examine the duality with the help of the matrix of the direct paths. It is known that the net model, based on the directed vertex graph can be constructed by any arbitrary given binary relation's matrix without any difficulties [6]. Some $L$ matrixes allow constructing the directed edge graph, but it can be done only occasionally. An example of the problem, when the construction of the vertex graph can be accomplished, and the construction of the edge graph can't be done, is presented in fig. 1.

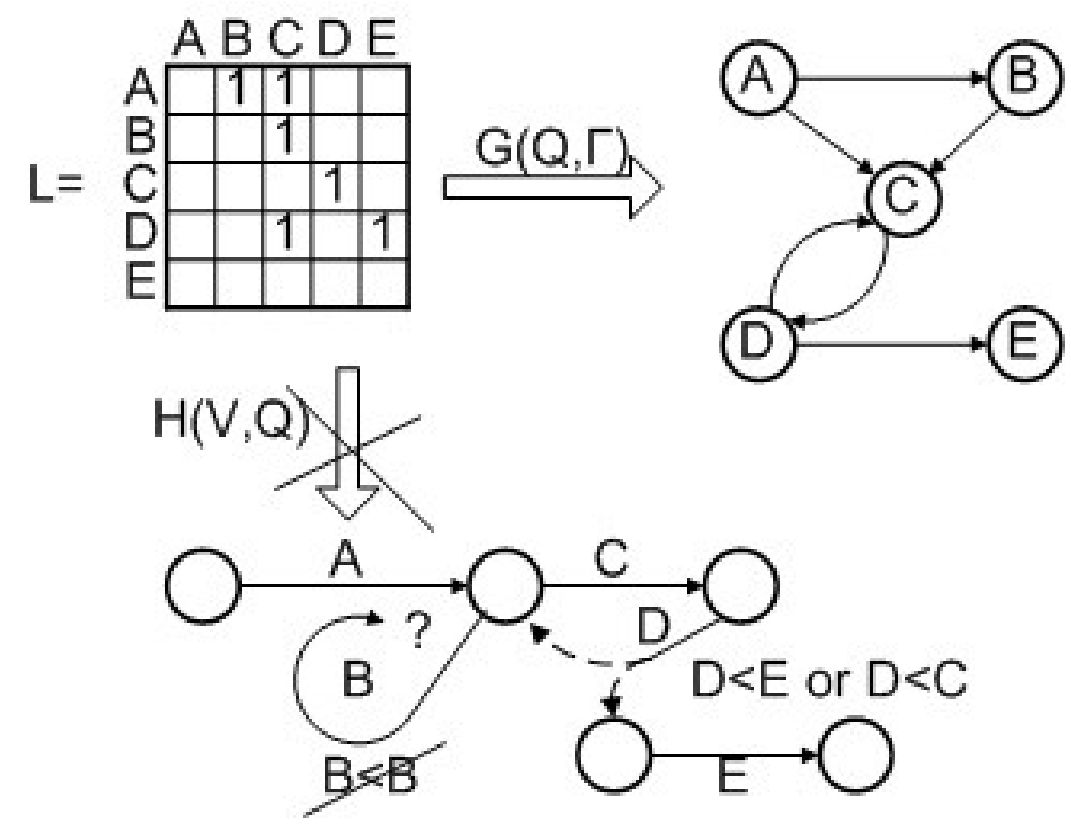

Fig. 1.

Indeed, let us try to join the end of the $B$ edge with the beginning of the $C$ edge. It will lead to the appearance of the $l_{BB} = 1$ loop, which is contradictory to the given $L$ matrix. We can try to join the end of the $D$ edge with the beginning of the $C$ edge. But on account of the $l_{DE} = 1$ condition the beginning of the $E$ edge must coincide with the end of the $A$ edge. This will lead to the appearance of the new $l_{AE} = 1$ element, and it is also contradictory to the given $L$ matrix. So, we can't always construct the edge graph by the arbitrary given $L$ matrix. It endorses the conclusion that the arbitrary given binary relation's matrix does not necessarily have the duality property. In other words, if we can always associate every row of the $L$ matrix with the $q_i$ vertex of the $G$ graph, it doesn't mean that we can associate every row of the $L$ matrix with the $q_i$ edge of the $H$ graph.

Another example may be adduced. Let us scrutinize the arbitrary given directed edge $H$ graph, which has not more than one initial and final edge (fig. 2).

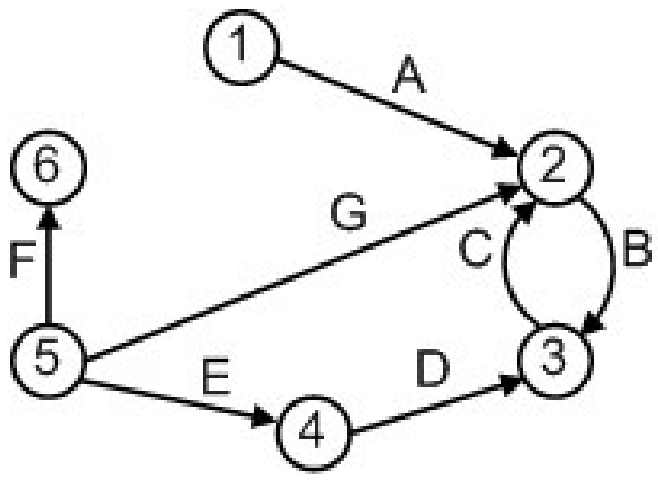

Fig. 2.

For this graph (already constructed) we can always arrange the edge's adjacency matrix (fig. 3a), which also allows constructing some vertex $G$ graph by it (fig. 3b). Matching both the $H$ and the $G$ graphs (fig. 4), it is easy to make sure that if both $q_i$ and $q_j$ edges are adjacent in the $H$ graph, then in the $G$ graph both $q_i$ and $q_j$ vertexes corresponding to them, are also adjacent [6].

$L = R =$

| | A | B | C | D | E | G | F |
|---|---|---|---|---|---|---|---|
| A | | 1 | | | | 1 | |
| B | | | 1 | | | | |
| C | | 1 | | | | 1 | |
| D | | | 1 | | | | |
| E | | | | 1 | | | |
| G | | | | | 1 | | 1 |
| F | | | | | | | |

a)

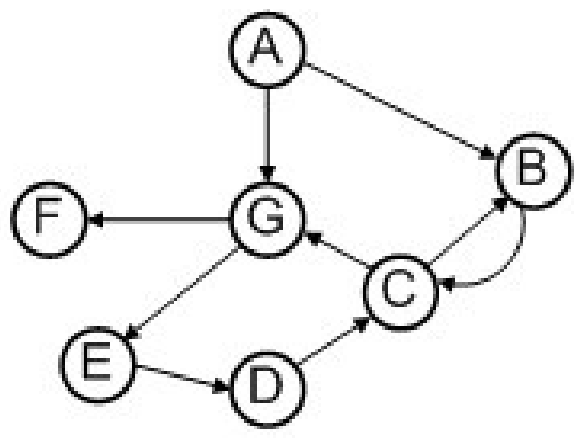

b)

Fig. 3.

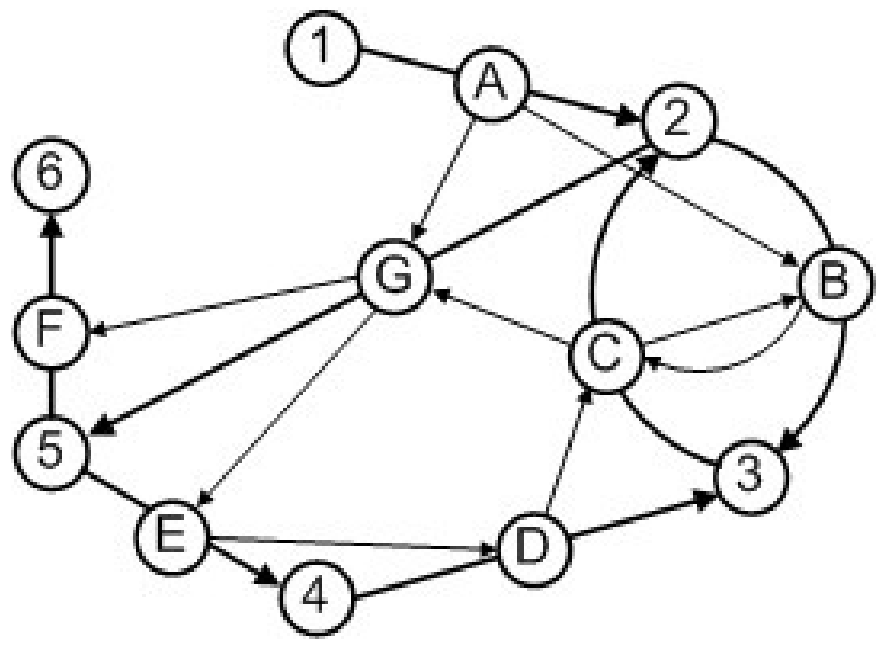

Fig. 4.

**Definition 1.1.** If the $L$ matrix at the same time appears to be the adjacency $E$ matrix of the vertex $G(Q,\Gamma)$ graph and also appears to be the adjacency $R$ matrix of the edge $H(V,Q)$ graph, then the vertex $G(Q,\Gamma)$ graph appears to be the adjacency graph of the edge $H(V,Q)$ graph's edges or, in other words, the adjacency graph of the $H(V,Q)$ graph. Thus, the concept of the duality for the directed graphs may be formulated this way: any arbitrary $H$ graph has the dual directed $G$ graph, which has such properties:

- Every edge in the $H$ graph corresponds to the vertex in the $G$ graph.
- The adjacent vertexes in the $G$ graph are the vertexes to which the adjacent between them edges in the $H$ graph correspond.

But from this definition it doesn't mean that any directed $G$ graph has the $H$ graph dual to it. On the contrary, from the analyzed examples a different conclusion follows: an arbitrary directed $G$ graph as a rule has not the dual directed $H$ graph. In other words, the duality property both as for the undirected graphs and for the directed graphs appears to be not obligatory reversible.

Sometimes the $G(Q,\Gamma)$ graph is called as a graph conjugated to the $H(V,Q)$ graph. This is based on the definition of the conjugated mapping. Suppose both $G(Q,\Gamma)$ and $H(V,Q)$ graphs are given. At that the $G(Q,\Gamma)$ graph appears to be the adjacent graph of the $H(V,Q)$ graph's edges. An incidence $S_G$ matrix of the edges [7, 8] of the $H(V,Q)$ graph defines the $f_G$ mapping of the $Q$ set onto the $\Gamma$ set: $f_G: Q \to \Gamma$. On the other hand the incidence $S_H$ matrix of the edges of the $H(V,Q)$ graph defines the $f_H$ mapping of the $V$ set onto the $\vec{Q}$ set. But then the $S_H^T$ matrix defines the $f_H^{-1}$ mapping of the $\vec{Q}$ set onto the $V$ set: $f_H^{-1}: \vec{Q} \to V$. Here and then let us agree to denote the set of the edges as $\vec{Q}$ if it is necessary to distinguish it from the set of the vertexs. In cases, when it does not make a mess, we'll denote them equally.

From theorems below it follows that if the single-valued mapping $\varphi: Q \to \vec{Q}$ exists, then in the given binary relations $L \subset Q \times Q$ system also exists the single-valued mapping $\psi: \Gamma \to V$. But then bicommutative diagram can be arranged [9], which is presented on fig. 5. The expression $f_G \circ \psi = \varphi \circ f_H^{-1}$ follows from the diagram. In this case we can say that both $f_G$ and $f_H^{-1}$ mappings are conjugated. In other words, the $f_G$ mapping is conjugated according to the $f_H^{-1}$ mapping.

Since the $f_G$ mapping defines the $G(Q,\Gamma)$ graph and the $f_H^{-1}$ mapping defines the $H(V,Q)$ graph, then the $G(Q,\Gamma)$ graph may be denoted as conjugated to the $H(V,Q)$ graph. The name of the $G(Q,\Gamma)$ graph may hereinafter define the name of the conjugated net models. From the analysis of the duality properties follows that the conjugated graph may be constructed not only by the initial information, but also by the edge graph. The edge graph as a rule couldn't be constructed by the vertex graph.

The advantages of the edge graphs compared to the vertex graphs become apparent also in the series of the theory graph's problems. The two classical problems exist: Euler problem on Seven Königsberg's bridges and Hamilton problem. Euler problem consists in finding the paths (directed in general), which take a route along the graph's edges. Hamilton problem includes finding of the paths (also directed in general), which take a route along the graph's vertexes.

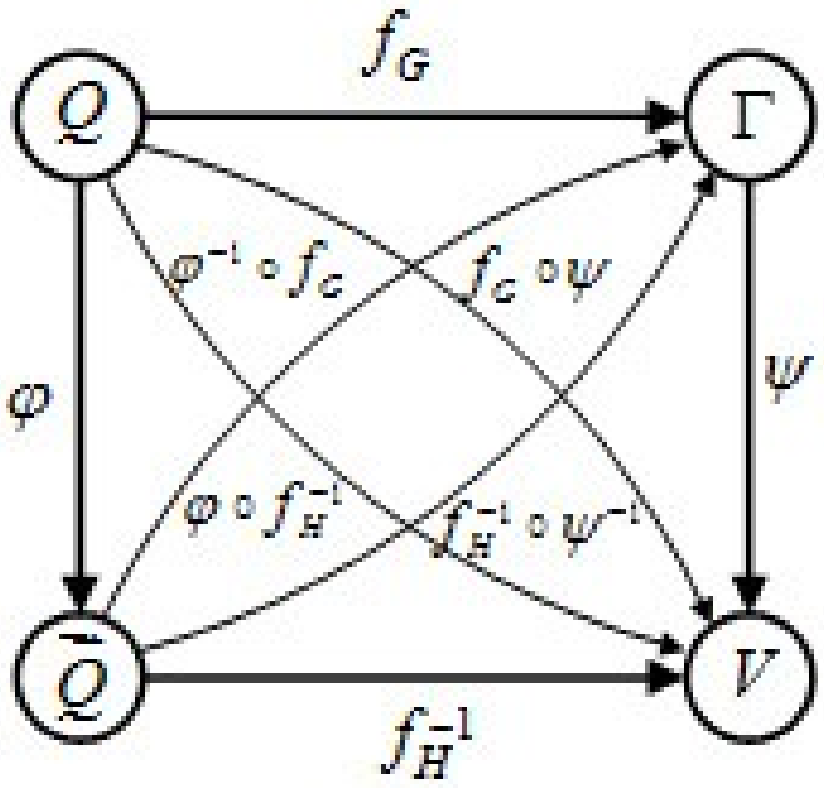

Fig. 5.

So, Euler problem may be suggested as the problem of ordering of the edges of the edge graph. Hamilton problem may be suggested as the problem of ordering of the vertexes of the vertex graph. In spite of the similarity in formulating, both problems have very little in common [2]. The theorems, which were proved by Euler, allow responding to a query of the existence of the Euler paths unambiguously by the graph's appearance. For the Hamilton paths we do not have such algorithms. For the most of the graphs we do not have even satisfactory algorithms, which can allow establishing the presence of the Hamilton paths.

The Dirak condition exists only for the undirected graphs [2]. It allows nothing more than dividing graphs into two classes: the graphs, in which at least one Hamilton cycle exists, can be referred to one class; the graphs, according to which no statement about existing of the Hamilton cycle in them can be expressed, can be referred to the other class. The only common regular approach in order to search Hamilton cycles nowadays is the method of the complete enumeration. But it is not convenient even if we have up-to-date computers.

As a result, we have the following contradiction. An analysis of the complicated process allows determining the list of simple elements and the binary relation's system on the set of these elements that is: to construct a matrix of the direct paths. A matrix of the direct paths in most cases allows constructing the directed vertex graph, the analysis of which or, in other words, the searching of the appropriate paths going through its vertexes, is coupled with such unconquerable difficulties, that it appears to be NP-complete.

A matrix of the direct paths in most cases does not allow constructing the edge graph, the analysis of which theoretically is always possible, and sometimes it can be very simple.

An arisen contradiction leads to the following target settings:

1. The establishment of both the possibility and the conditions of the existence of such direct path's (binary relations) matrix, which always can be accepted as the adjacency matrix of both the vertexes of the vertex graph and the edges of the edge graph.

2. The establishment of both the possibility and the conditions of bringing to such dual form any direct path's matrix.

3. If such matrix exists, then both algorithms and examples of the possibility of their usage in order to construct the pictorial representations of net models of complicated both processes and systems should be constructed.

Finally, we are searching for a possibility of constructing an edge graph according to the given vertex graph.

## 2 The quasicanonical vertex and edge graphs

An arbitrary compound process can be presented as a net model: either ordinary or conjugated [10]. They correspond to either vertex or edge graphs, which are dual. A structural similarity demands the equality of the following topological invariants [10]:

1. The numbers of the elements of both the initial and the fundamental sets.
2. The binary relation's systems, assigned on the fundamental sets of the elements.
3. The cyclomatic numbers (but not in all cases).

Both graphs must be restricted, so they must not have the contours. Taking up the graphs with the contours does not affect the demands of the structural similarity, because the contours are the tool for the decreasing both the model's and the adjacency matrix's size [10]. It is necessary for the number of the reiterations in the contours to be both finite and equal to the number of the reiterations in the contours of the compound process for the retaining of the structural similarity properties.

A model, as well as the compound process, must possess the property of the insularity: every boundary element of the real process must correspond to the boundary element of the model. On account of this claim both the initial and the final graph's vertexes must be included in the number of the model's elements, though they do not reverberate obviously, as other vertexes, in the adjacency matrix.

Meanwhile we'll examine only the variant of the quasiduality (the incomplete duality) between both vertex and edge graphs. In this case the graphs may have the different cyclomatic numbers. Other criteria of the structural similarity must be equal.

### Theorem 1 "On a quasicanonical adjacency matrix"

A theorem on the quasicanonical adjacency matrix determines both necessary and sufficient conditions that the direct path's $L$ matrix has a dual nature that is at the same time it might be both $E-$ adjacency matrix of the $G$ graph's vertexes and $R$ − adjacency matrix of the $H$ graph's edges on condition that they may have not equal cyclomatic numbers.

Let's prove the theorem for the case of the directed graphs, since any undirected $G$ graph may be transformed into a directed one by the operation of the redoubling [2].

It is given: a set $Q = \{q_i\}$ and $L = Q \times Q$ by the way of $q_i < q_j$, and $L = \|l_{ij}\|_1^n = \|e_{ij}\|_1^n$ for the $G(Q, \Gamma)$ graph. Then $\|l_{ij}\|_1^n = \|r_{ij}\|_1^n = R$ for the $H(Q, V)$ graph if and only if:

**Condition 1**

$$\left.\begin{array}{l} 1.\ \text{An } L = \|l_{ij}\|_1^n \text{ generates } C_n = \|c_{ij}\|_1^n = [0] \\ 2.\ A \text{ minor } |l_{ij}|_1^{n-1} \text{ of every } l_{ij} = 1 \text{ generates } C_{n-1} = \|c_{ij}\|_1^{n-1} = [0] \end{array}\right\}$$

Where:

$c_{ij} = l_{ij}(\Delta_{j/i}s_{ij} + \Delta_{i/j}s_{ij})$;

$\Delta_{j/i}s_{ij} = (s_{ij} - \min_j s_{ij})_i$;

$\Delta_{i/j}s_{ij} = (s_{ij} - \min_i s_{ij})_j$;

$s_{ij} = l_{ij} * (\sum_{\substack{i=1 \\ j}}^{k} l_{ij} + \sum_{\substack{j=1 \\ i}}^{k} l_{ij})$;

$\left(\min_j s_{ij}\right)_i = \min_{j/i} s_{ij} \in \{s_{ij} \neq 0\}$;

$\left(\min_i s_{ij}\right)_j = \min_{i/j} s_{ij} \in \{s_{ij} \neq 0\}$;

$k = n, (n-1)$;

For the proof of the theorem it must be testified that the $L$ matrix, which meets the condition (1) and is considered as the $R$ matrix – the adjacency matrix of $H$ graph's edges, has all the information for a single-valued representation of the $F$ matrix – the adjacency matrix of $H$ graph's vertexes.

**Proof**

It is given: a set $Q = \{q_i\}$ and a $\|l_{ij}\|_1^n = L = Q \times Q$ matrix (fig.6).

Let us examine this matrix as matrix $L$ − the adjacency matrix of the $G$ graph's vertexes and at the same time as matrix $R$ – the adjacency matrix of the $H$ graph's edges.

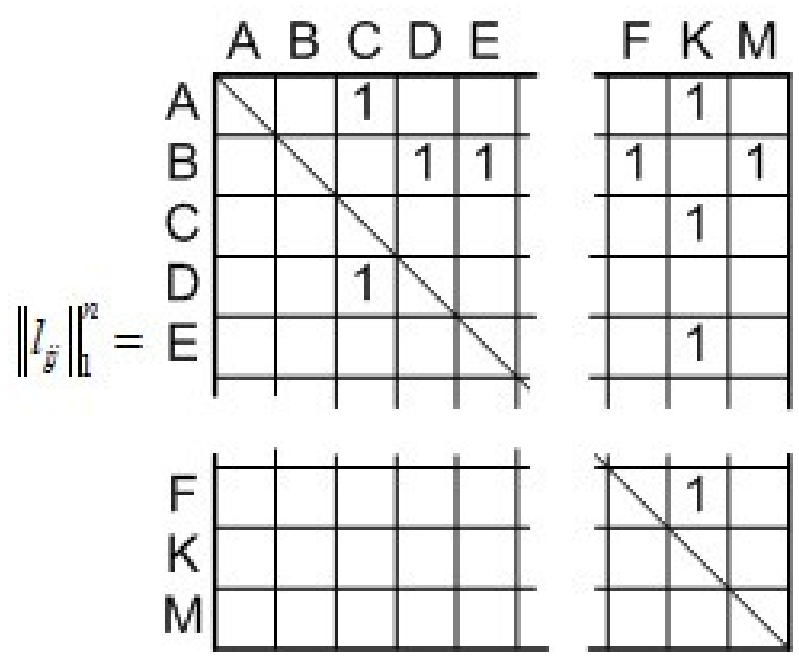

Fig. 6. A $\boldsymbol{L} = \boldsymbol{Q} \times \boldsymbol{Q}$ matrix

Let us prove that in order to construct both $G$ and $H$ graphs by the $L$ matrix, it is necessary and sufficient that the $L$ matrix could be disintegrated naturally into the $\left|l_{ij}\right|_k^p$ submatrixes, which have the size of $(k \times p)_h, (h = 1,2, \dots, m-2)$, all the elements of which are units, and all other elements of the $L$ matrix, which are not part of such $\left|l_{ij}\right|_k^p$ submatrixes – zeros. Let's prove that the $L$ matrix consists of such $\left|l_{ij}\right|_k^p$ submatrixes if and only if the condition (1) is fulfilled.

**Definition 2.1.** We'll denote either the $G_{i,j}$ subgraph or the $H_{i,j}$ subgraph, which corresponds to either the $i$ line or the $j$ column of the $L$ matrix as a fragment of either $G$ or $H$ graphs.

Let's choose both an arbitrary $i = B$ line and an arbitrary $j = K$ column. In order to construct a fragment of the $G$ graph by the $L$ matrix (fig.6), which will correspond to the line $i = B$ (fig. 7a), it is enough to draw the $q_B$; $q_D$; $q_E$; …; $q_M$ vertexes and connect the vertex $q_B$ with the $q_D$; $q_E$; …; $q_F$; $q_M$ vertexes by the edges. In a similar manner we can construct a fragment of the $G$ graph, which corresponds to the column $j = K$ (fig. 7b). Getting across one line to another it is possible to construct the whole $G$ graph.

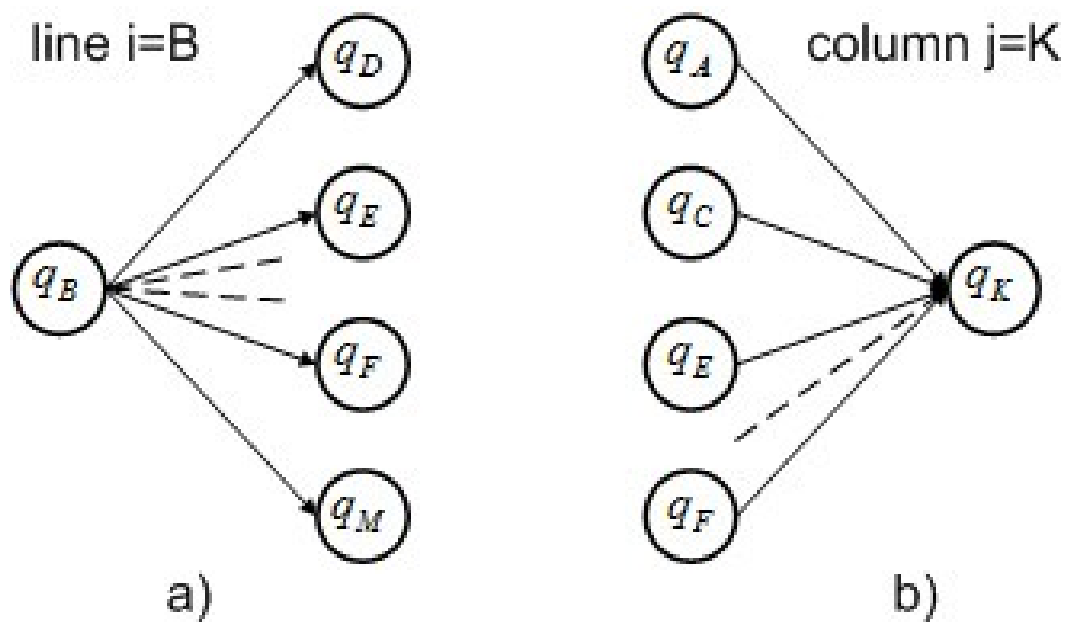

Fig. 7. a) A fragment of the $\boldsymbol{G}$ graph, which corresponds to the $\boldsymbol{i} = \boldsymbol{B}$ line;
b) A fragment of the $\boldsymbol{G}$ graph, which corresponds to the $\boldsymbol{j} = \boldsymbol{K}$ column

A construction of the $H$ graph's fragments is carried out differently. According to the definition, if $l_{ij} = r_{ij} = 1$, then the $q_i$ edge ends in the same vertex, in which the $q_j$ edge begins. Thus, if in the $i = B$ line the elements $l_{BD} = l_{BE} = \dots = l_{BM} = 1$, then the $q_B$ edge finishes in the same $v_h$ vertex, in which the edges $q_D$; $q_E$; …; $q_F$; $q_M$ begin. A $v_h$ vertex hasn't an explicit view in the $\left\|l_{ij}\right\|_1^n$ matrix, yet it is determined by all the $l_{ij} = 1$ elements in the $i = B$ line. It is obvious that in order to construct a fragment $H_{B,j}(V, Q)$ of the $H(V, Q)$ graph it is enough to draw the $v_h$ vertex and the $q_D$; $q_E$; …; $q_F$; $q_M$ edges in such a way that the $q_B$ edge will finish in the $v_h$ vertex while the $q_D$; $q_E$; …; $q_F$; $q_M$ edges will begin in the $v_h$ vertex (fig 8a). A $H$ graph's fragment, which is corresponding to the $j = K$ column, may be constructed in a similar manner (fig.8b). At that the $v_h$ vertex will be determined by the all $l_{ij} = 1$ elements, which enter the $j = K$ column. However, it is obvious, that as a rule you can not construct the $H$ graph from such fragments.

Let us examine the rules of constructing the $H$ graph from such fragments and prove that both the two $q_{i_1}$ and $q_{i_2}$ edges may finish in one $v_h$ vertex if and only if both the $i_1$ and $i_2$ lines of the $\left\| l_{ij} \right\|_1^n$ matrix are congruent by all the $j$.

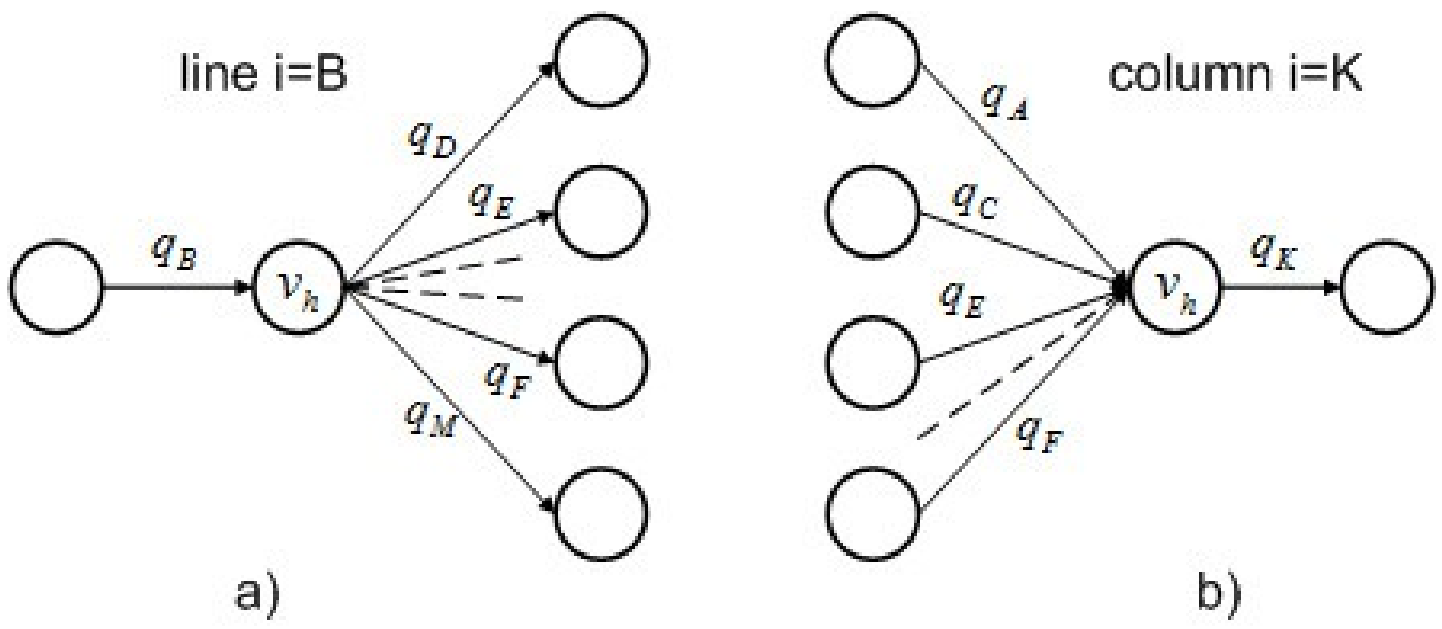

Fig. 8. A construction of the ***H*** graph's fragments

Let $l_{i_1 j} = 1$ at $j = k, l, \dots, p$ and $l_{i_2 j} = 1$ at $j = r, s, \dots, u$, and let the $q_{i_1}$ and $q_{i_2}$ edges end in one $v_h$ vertex. Then from the $q_{i_1}$ edge's end the $q_k, q_l, \dots, q_p$ edges come out, and from the $q_{i_2}$ edge's end – $q_r, q_s, \dots, q_u$ edges come out. Since as both $q_{i_1}$ and $q_{i_2}$ edges are supposed to end in the $v_h$ vertex, then every $q_j^{(+)}(v_h)$ edge, coming out of the $v_h$ vertex, belongs to either the $\{q_k, q_l, \dots, q_p\}$ set or to the $\{q_r, q_s, \dots, q_u\}$ set, or to both of them, that is:

$$\left\{q_j^{(+)}(v_h)\right\} = \{q_k, q_l, \dots, q_p\} \cup \{q_r, q_s, \dots, q_u\}$$

On the other hand, every $q_j^{(+)}(v_h)$ edge, which comes from the $v_h$ vertex, at the same time, comes from both $q_{i_1}$ edge's end and the $q_{i_2}$ edge's end, in other words, it simultaneously belongs to both the $\{q_k, q_l, \dots, q_p\}$ set and the $\{q_r, q_s, \dots, q_u\}$ set, or it equivalently belongs to the intersection of these sets. Then:

$$\left\{q_j^{(+)}(v_h)\right\} = \{q_k, q_l, \dots, q_p\} \cap \{q_r, q_s, \dots, q_u\}$$

From premises follows:

$$\{q_k, q_l, \dots, q_p\} \cup \{q_r, q_s, \dots, q_u\} \equiv \{q_k, q_l, \dots, q_p\} \cap \{q_r, q_s, \dots, q_u\}$$

And this is possible only in case, if:

$$\{q_k, q_l, \dots, q_p\} \equiv \{q_r, q_s, \dots, q_u\}$$

As a result both $i_1$ and $i_2$ lines must be congruent by all $j$. Thereby we came to the following condition. The both $q_{i_1}$ and $q_{i_2}$ edges end in one $v_h$ vertex and only if both the $i_1$ and $i_2$ lines of the $L$ matrix, corresponding to them, are congruent by all $j$. It becomes evident that the premises condition is right not only for the two, but for any number of edges, which end in one $v_h$ vertex.

By analogy we come to the next condition. Similarly we can prove that the $q_{j_1}, q_{j_2}, q_{j_3}, \dots, q_{j_p}$ edges begin in the same $v_h$ vertex if and only if the corresponding $j_1, j_2, j_3, \dots, j_p$ columns of the $L$ matrix are congruent by all $i$. The premises conditions are combined evidently in the final condition:

**Condition 2**

The $H$ graph's $q_{i_1}, q_{i_2}, q_{i_3}, \dots, q_{i_k}$ edges end, and the $q_{j_1}, q_{j_2}, q_{j_3}, \dots, q_{j_p}$ edges begin in the one and the same $H$ graph's $v_h$ vertex if and only if from the $l_{ij} = 1$ elements $\left(i = i_1, i_2, i_3, \dots, i_k;\ j = j_1, j_2, j_3, \dots, j_p\right)$ we can construct the $\left|l_{ij}\right|_k^p$ submatrix with the $(k \times p)_h$ size, and at that time $l_{ij} = 0$ for all $i \neq i_1, i_2, i_3, \dots, i_k$ at $j = j_1, j_2, j_3, \dots, j_p$ and for all $j \neq j_1, j_2, j_3, \dots, j_p$ at $i = i_1, i_2, i_3, \dots, i_k$.

Consequently, if the $L$ matrix appears to be the adjacency $R$ matrix of the $H$ graph's edges, then for every $H$ graph's vertex it is possible to set up a correspondence to the $\left|r_{ij}\right|_k^p = \left|l_{ij}\right|_k^p$ submatrix, which size is $(k \times p)_h$, of the $R = L$ matrix, which is made up of the $r_{ij} = l_{ij} = 1$ elements. All the $r_{ij} = l_{ij}$ elements, which do not belong to such submatrixes, are equal to zero. Both the initial and the final $H(V, Q)$ graph's vertexes are the exclusion. An initial vertex corresponds to an empty column, and the final vertex – to an empty line.

This condition is both necessary and sufficient in order that the $L$ matrix could be the $R$ matrix – the adjacency matrix of the $H$ graph's edges. The first part of the proof is fulfilled.

Let's turn to the second part of the proof and make it evident that the $\left\|l_{ij}\right\|_1^n$ matrix satisfies to the condition (2) if and only if it satisfies to the condition (1).

Let's mark that the $r_{ij} = l_{ij} = 1$ elements, which do not correspond to the $H(V, Q)$ graph's vertexes, couldn't be present in the $L = R$ matrix.

**Definition 2.2.** The number of the edges, which go out of the $v_h$ vertex is named the out-degree of the $v_h$ vertex and have the sign (+), the number of the edges, which enter the $v_h$ vertex is named the in-degree of the $v_h$ vertex and have the sign (-) [7, 8]. We'll take in account only the modulus of such values.

It is evident that $\sum_{\underline{j=1}\atop i_1}^{n} r_{ij}$ – is the out-degree of the $G$ graph's $q_{i_1}$ vertex and at that time it is the out-degree of the $v_h$ vertex, in which the $q_{i_1}$ edge is ending and the $H$ graph's $q_j$ $(j = j_1, j_2, j_3, \dots, j_p)$ edges (at the condition that we'll gain in constructing the $H$ graph) are beginning.

On the other hand, $\sum_{\underline{i=1}\atop j_1}^{n} r_{ij}$ – is the in-degree of the $G$ graph's $q_{i_1}$ vertex and at the same time it is the in-degree of the $v_h$ vertex in which the edges $q_i$ $(i = i_1, i_2, i_3, \dots, i_k)$ are ending and the $H$ graph's $q_{j_1}$ edge is beginning (at the condition that we'll gain in constructing the $H$ graph).

Therefore (from condition 2):

$\sum_{\underline{j=1}\atop i_1}^{n} r_{ij} = P_h$ at $i = i_1, i_2, i_3, \dots, i_k$

$\sum_{\frac{i=1}{j_1}}^{n} r_{ij} = K_h$ at $j = j_1, j_2, j_3, \dots, j_p$

Then the sum of both the in-degrees and out-degrees of the $v_h$ vertex will be: $S_h = K_h + P_h$

A complete number of the connections (binary relations), which correspond to the $v_h$ vertex is equal to the product: $K_h * P_h$.

Let us examine an arbitrary submatrix, which contains units. A $L = R$ matrix is presented in fig. 9. A similar $\left|r_{ij}\right|_k^p$ submatrix corresponds to some $v_h$ vertex.

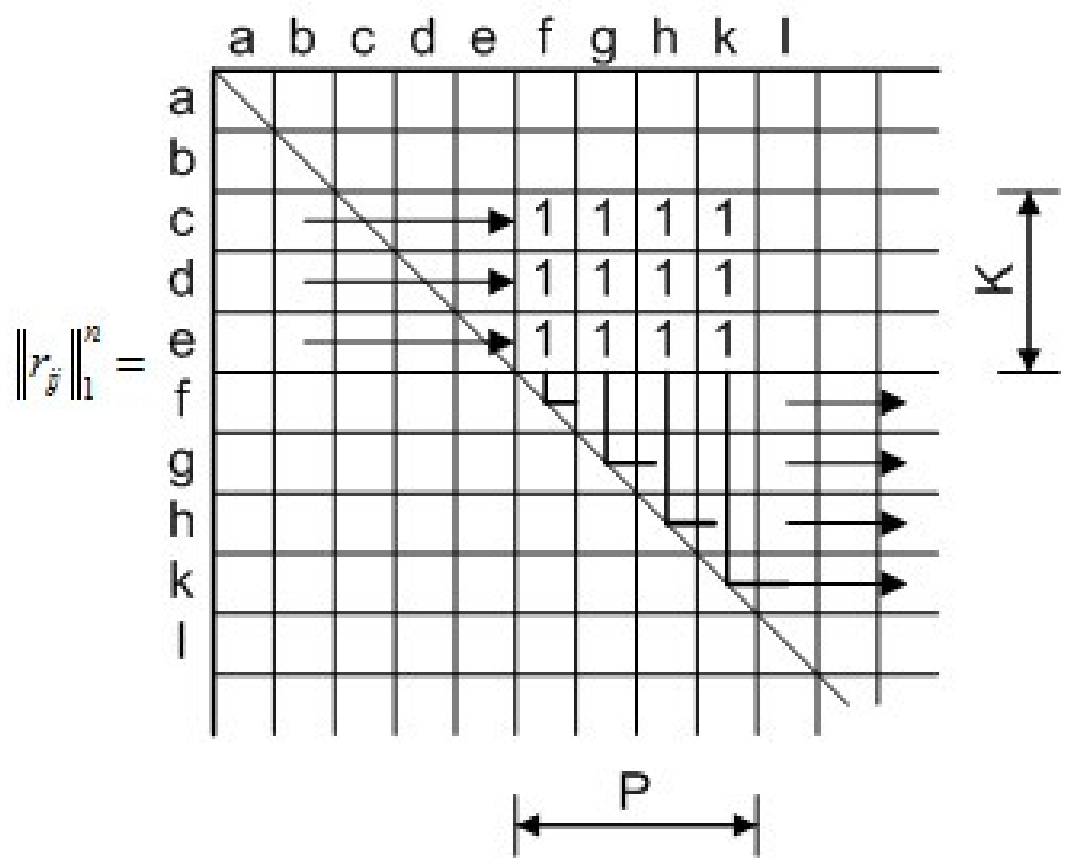

Fig. 9.

Let's represent the $v_h$ vertex in a view of the circle (fig. 10). On the left side we'll mark the points of adding the endings of the incoming edges, on the right side – the points of adding the beginnings of the outgoing edges, and connect the received points with the arrows in compliance with the $\left|r_{ij}\right|_k^p$ submatrix. It is evident that the $\left|r_{ij}\right|_k^p$ submatrix entirely corresponds to the $v_h$ vertex.

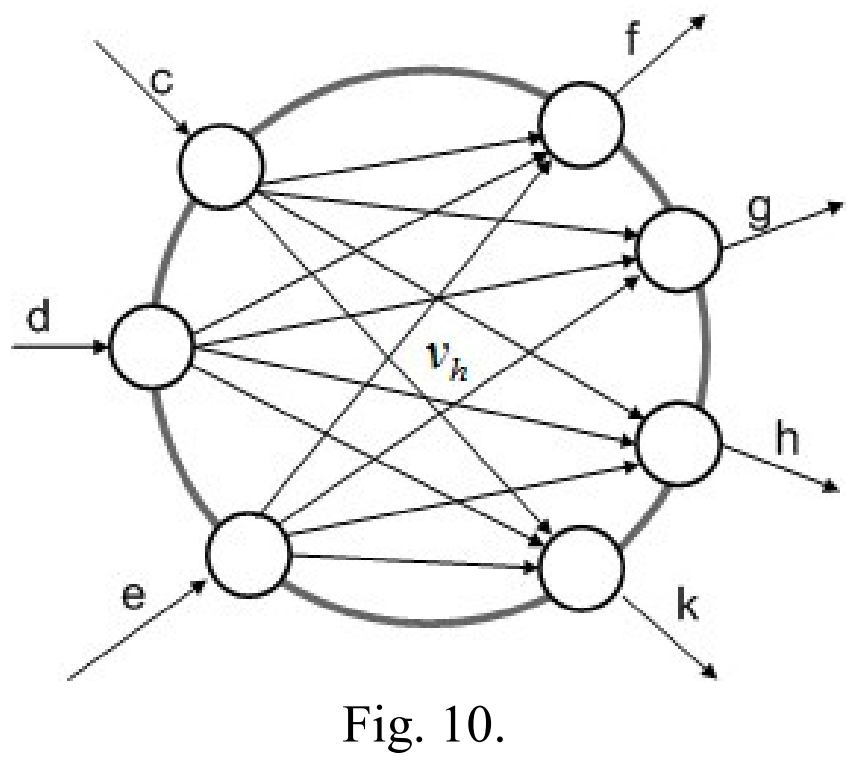

Fig. 10.

Let's examine any of these connections: for instance, the arrow "$cf$". There are the $p$ connections which go out from the point "$c$", and the $k$ connections which come into the point "$f$". So, the $r_{cf} = 1$ element has the $(P_h - 1)$ going out connections, parallel to it and the $(K_h - 1)$ going into connections, also parallel to it. We can say that $(P_h + K_h - 2)$ is the total complexity of all the connections, parallel to the chosen element. The complete number of the connections (binary relations), which can characterize every $r_{ij} = 1$ element, being part of the $(k \times p)_h$ submatrix (the $r_{ij} = 1$ element and the connections, parallel to it, are included), will be: $S_h - 1 = K_h + P_h - 1$.

So, every element $r_{ij} = 1$ corresponding to the $v_h$ vertex may be characterized by the sum: $S_h = K_h + P_h$, which is per unit more then total number of all connections in the $v_h$ vertex. If the element is the $r_{ij} = 0$, then of course it is corresponding to none of the vertexes and then the number of the connections, parallel to it, ought to be equal to zero.

Using the $S_h$ value, let's try to work out the quantitative assessment of the fact whether the $L$ matrix can be disintegrated into the $\left|l_{ij}\right|_k^p$ submatrixes or not. It will help to answer the question: whether the $L$ matrix can be at the same time the $R$ matrix. First of all for this purpose let's agree to characterize every $L$ matrix's $l_{ij} = 1$ element by the value:

$$s_{ij} = s_h = l_{ij}\left(\sum_{\substack{i=1\\j}}^{n} l_{ij} + \sum_{\substack{j=1\\i}}^{n} l_{ij}\right)$$

Also we'll agree to characterize the $L$ matrix on the whole with the corresponding $\left\|s_{ij}\right\|_1^n$ matrix. It is evident that for the $l_{ij} = 1$ elements, which are part of the $\left|l_{ij}\right|_k^p$ submatrix, a condition is right: for all the $\left(l_{ij} = 1\right) \in v_h$ elements – $s_{ij} = s_h = const$.

Otherwise the $l_{ij} = 1$ elements could not belong to one $v_h$ vertex.

Let us agree to characterize every $l_{ij} = 1$ element with one more value: the sum of the excess values of both in-degree's and out-degree's sums for the element in question in comparison with such elements, which have the minimum values of such sums in the corresponding line and column.

The sum of the excess values of both in-degree's and out-degree's sums may be determined as:

$$c_{ij} = l_{ij}\left(\Delta_{\frac{j}{i}} s_{ij} + \Delta_{\frac{i}{j}} s_{ij}\right)$$

Where:

$$\Delta_{\frac{j}{i}} s_{ij} = \left(s_{ij} - \min_j s_{ij}\right)_i$$

$$\Delta_{\frac{i}{j}} s_{ij} = \left(s_{ij} - \min_i s_{ij}\right)_j$$

In the formulas above:

$s_{ij}$ - can be determined by premisis formula;

$\left(\min_j s_{ij}\right)_i = \min_{j/i} s_{ij} \in \left\{s_{ij} \neq 0\right\}$ – is the minimum value of both in-degree's and out-degree's sums for the $l_{ij} = 1$ elements at $j$ in the $i$ line;

$\left(\min_i s_{ij}\right)_j = \min_{i/j} s_{ij} \in \left\{s_{ij} \neq 0\right\}$ – is the minimum value of both in-degree's and out-degree's sums for the $l_{ij} = 1$ elements at $i$ in the $j$ column;

$\Delta_{\frac{j}{i}} s_{ij}$ – an excess of the value of both in-degree's and out-degree's sums for the given $l_{ij} = 1$ element in the $i$ line in comparison with the element, which has in this line the minimum value: $s_{ij} \neq 0$;

$\Delta_{\frac{i}{j}} s_{ij}$ – an excess of the value of both in-degree's and out-degree's sums for the given $l_{ij} = 1$ element in the $j$ column in comparison with the element, which has in this line the minimum value $s_{ij} \neq 0$;

Let us examine some computational exercises on the $c_{ij}$ value. The arbitrary $L$ matrix is given on fig. 11. Also we have calculated both $\sum_{\substack{j=1 \\ i}}^{n} l_{ij}$ and $\sum_{\substack{i=1 \\ j}}^{n} l_{ij}$ values. A $\|l_{ij}\|_1^n$ matrix, which elements are calculated by previous formula, is represented in fig. 11. We can also calculate the elements of the $\|s_{ij}\|_1^n$ matrix and the $\|c_{ij}\|_1^n$ matrix's elements by the premisis formula (fig. 12a, b).

$\|l_{ij}\|_1^n =$

| | a | b | c | d | e | f | g | h | $\sum_{\substack{j=1 \\ i}}^{n} l_{ij}$ |
|---|---|---|---|---|---|---|---|---|---|
| a | | 1 | 1 | 1 | | | | 1 | 4 |
| b | | | 1 | | 1 | | | 1 | 3 |
| c | | | | | | 1 | 1 | | 2 |
| d | | | | | | 1 | 1 | | 2 |
| e | | | | | | | | 1 | 1 |
| f | | | | | 1 | | | 1 | 2 |
| g | | | | | 1 | | | | 1 |
| h | | | | | | | | | 0 |
| $\sum_{\substack{i=1 \\ j}}^{n} l_{ij}$ | 0 | 1 | 2 | 1 | 3 | 2 | 2 | 4 | |

Fig. 11. An arbitrary ***L*** matrix

The calculations are done the following way:

1. $s_{be} = 6$
2. $\left(\min_j s_{bj}\right)_{i=b} = s_{bc} = 5$
3. $(\min_i s_{ie})_{j=e} = s_{ge} = 4$
4. $\Delta_{\frac{j}{i=b}} s_{bj} = 6 - 5 = 1$
5. $\Delta_{\frac{i}{j=e}} s_{ie} = 6 - 4 = 2$
6. $c_{be} = 1 * (1 + 2) = 3$

A vertex $G(Q, \Gamma)$ graph corresponding to the given $L$ matrix is represented in fig. 13. Both the $s_{ij}$ values (in the numerator) and the $c_{ij}$ (in the denominator) are indicated close to the graph's edges.

We can make use of the quantitative assessment of the $L$ matrix's elements in order to find the necessary conditions for the $L$ matrix to be the $R$ matrix. In other words, we must find the conditions of the $L$ matrix's disintegration into the $\left|r_{ij}\right|_k^p$ submatrixes.

a) $\left\|s_{ij}\right\|_1^n =$

| | a | b | c | d | e | f | g | h |
|---|---|---|---|---|---|---|---|---|
| a | \ | 5 | 6 | 5 | | | | 8 |
| b | | \ | 5 | | 6 | | | 7 |
| c | | | \ | | | 4 | 4 | |
| d | | | | \ | | 4 | 4 | |
| e | | | | | \ | | | 5 |
| f | | | | | 5 | \ | | 6 |
| g | | | | | 4 | | \ | |
| h | | | | | | | | \ |

b) $\left\|c_{ij}\right\|_1^n =$

| | a | b | c | d | e | f | g | h |
|---|---|---|---|---|---|---|---|---|
| a | \ | 0 | 2 | 0 | | | | 6 |
| b | | \ | 0 | | 3 | | | 4 |
| c | | | \ | | | 0 | 0 | |
| d | | | | \ | | 0 | 0 | |
| e | | | | | \ | | | 0 |
| f | | | | | 1 | \ | | 2 |
| g | | | | | 0 | | \ | |
| h | | | | | | | | \ |

Fig. 12. A calculation of the elements of both the $\left\|\boldsymbol{s_{ij}}\right\|_{\mathbf{1}}^{\boldsymbol{n}}$ matrix and the $\left\|\boldsymbol{c_{ij}}\right\|_{\mathbf{1}}^{\boldsymbol{n}}$ matrix

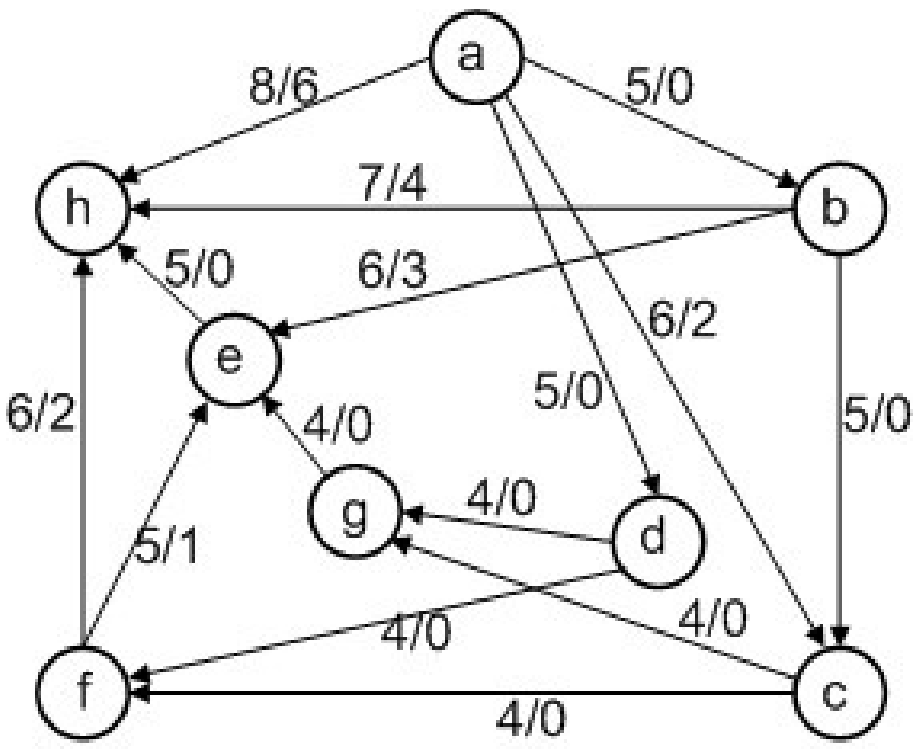

Fig. 13. A $\boldsymbol{G}(\boldsymbol{Q},\boldsymbol{\Gamma})$ vertex graph corresponding to the given $\boldsymbol{L}$ matrix

Let us examine the $\left|l_{ij}\right|_k^p = \left|r_{ij}\right|_k^p$ submatrix, which consists of units. It is corresponding to some $v_h$ vertex of the $H$ graph (fig. 9). It is obvious that all the $c_{ij}$ elements, corresponding to these submatrix's $l_{ij} = r_{ij} = 1$ elements, are equal to zero. So, if the $\left\|l_{ij}\right\|_1^n$ matrix can be disintegrated into the $\left|l_{ij}\right|_k^p$ submatrixes, all elements of which are: $l_{ij} = 1$, and all the $l_{ij}$ elements, which are outside such $\left|l_{ij}\right|_k^p$ submatrixes, are equal to zero, then matrix $\left\|c_{ij}\right\|_1^n = [0]$.

Let's assume that there are zeros among the $l_{ij}$ elements which belong to some $\left|l_{ij}\right|_k^p$ submatrix. Their quantity may vary in different lines and columns. Let's show that in this case at least one of the elements, corresponding to the $\left|l_{ij}\right|_k^p$ submatrix, is $c_{ij} \neq 0$. For this purpose let us examine an arbitrary $i^*$ line of such $\left|l_{ij}\right|_k^p$ submatrix (fig. 14). As the number of zeros in the various columns of the submatrix is different, it is obvious that the $s_{i^*j}$ values along the $i^*$ line will be different, so one of them will be marginal (zero values are excluded). Then, certainly, at least, for one of the $l_{i^*j} = 1$ elements, for instance, for one $i^*$ line we'll have: $\Delta_{\frac{j}{i^*}} s_{i^*j} \neq 0$ and $c_{i^*j} \neq 0$. Similarly we can achieve that at least one of the $l_{ij^*}$ elements at least for one of the $j^*$ columns in the $\left|l_{ij}\right|_k^p$ submatrix generates $c_{ij^*} \neq 0$, that is: $\Delta_{\frac{i}{j^*}} s_{ij^*} \neq 0$ and $c_{ij^*} \neq 0$.

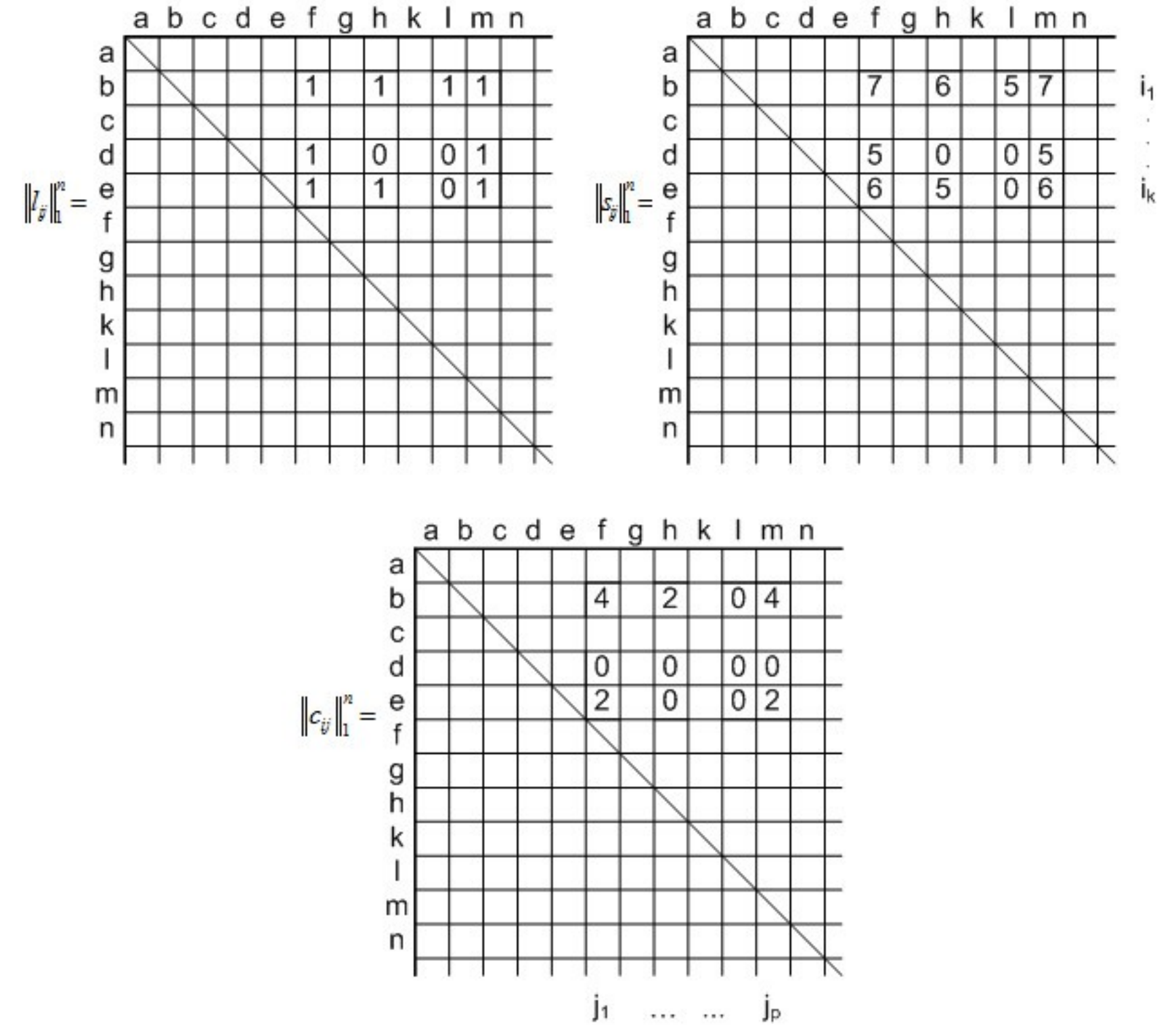

Fig. 14.

So, if among the elements, which arrange the $\left|l_{ij}\right|_k^p$ submatrix, there are zero elements and their numbers vary in different lines and columns, then, at least, one of the elements, corresponding to such $\left|l_{ij}\right|_k^p$ submatrix, is not equal to zero: $c_{ij} \neq 0$.

Thus, in order to the $l_{ij} = 1$ elements of the $\left|l_{ij}\right|_k^p$ submatrix enter the $H$ graph's $v_h$ vertex, each of the $l_{ij} = 1$ elements of the $\left|l_{ij}\right|_k^p$ submatrix, which correspond to the $v_h$ vertex, must satisfy to the following condition: $c_{ij} = 0$

Well, let us formulate the necessary condition.

In order to accept the $L$ matrix as the $R$ matrix it is necessary for the $L$ matrix to generate the $\left\|c_{ij}\right\|_1^n = [0]$ matrix. But the condition $\left\|c_{ij}\right\|_1^n = [0]$ is not sufficient to accept the $L$ matrix as the $R$ matrix.

Let's turn to the formulating of the sufficient condition. Let us suggest that there may be cases, when in the $\left|l_{ij}\right|_k^p$ submatrix part of elements $l_{ij} = 0$ (i.e. not all the elements will be equal to zero) and, at the same time, all $c_{ij} = 0$. An example of such vertex is presented in fig. 15. Part of the connections inside the vertex is absent. A question arises: does such a submatrix define the $v_h$ vertex?

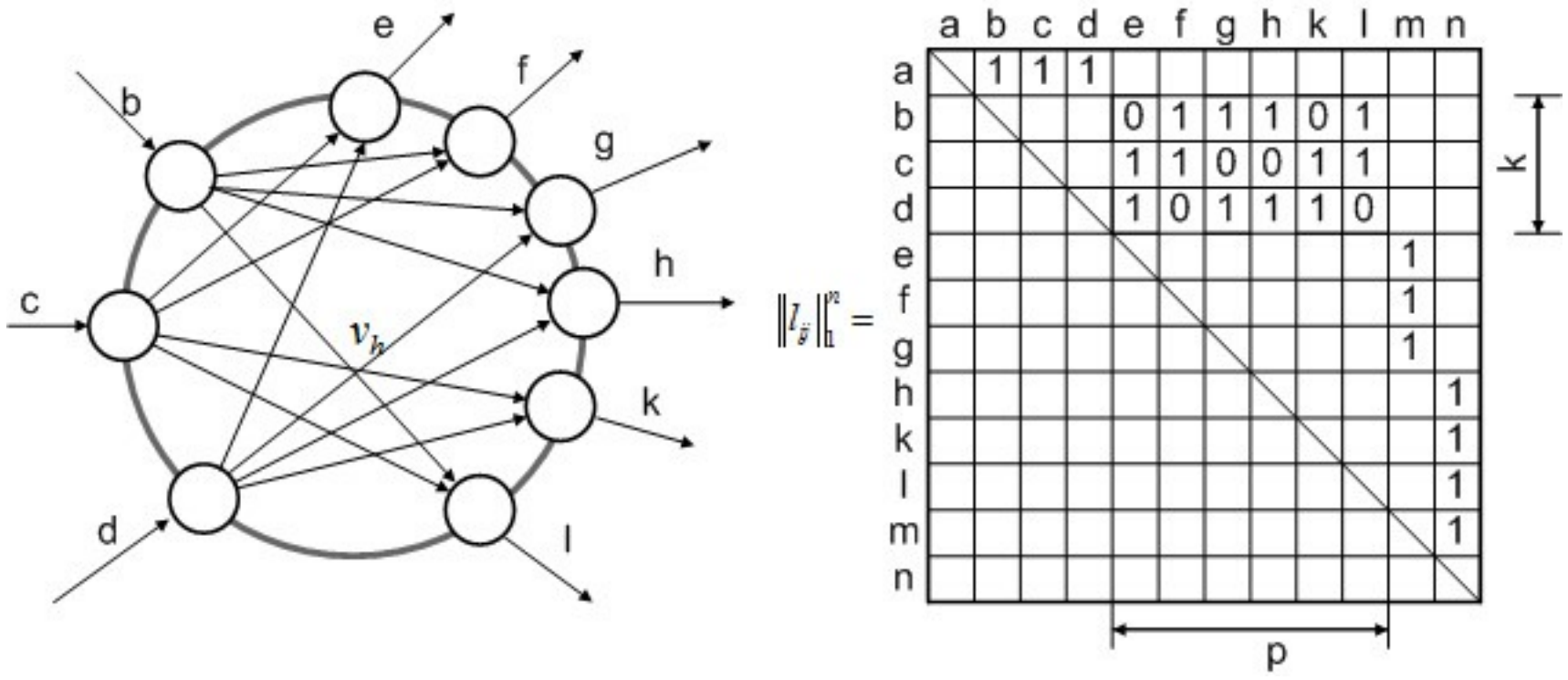

Fig. 15.

In this case the $\left|l_{ij}\right|_k^p$ submatrix must have such characteristics: there must be $k - t \neq 0$ of the $l_{ij} = 1$ elements in all of its columns, and in all the lines there must be $p - s \neq 0$ of the $l_{ij} = 1$ elements. In such a submatrix in any line and column there must be at least one $l_{ij} = 0$ element. By convention let us name such submatrix as dummy. It is evident that for all of the dummy submatrix's elements $c_{ij} = 0$, but at the same time such a submatrix defines no $v_h$ vertex of the $H$ graph. Let us permit that the dummy $\left|l_{ij}\right|_k^p$ submatrix is corresponding to some $v_h$ vertex and show that this permission is incorrect. From the previous it follows that if the all $l_{ij}$ elements of the $\left|l_{ij}\right|_k^p$ submatrix, which determine some $v_h$ vertex, are units, than the all corresponding to them $c_{ij} = 0$, and so the condition is right: $D_{k,p} = \sum_{i_1}^{i_k} \sum_{j_1}^{j_p} c_{ij} = 0$

Where: $i = i_1;\ i_2;\ i_3; \dots;\ i_k$, $j = j_1;\ j_2;\ j_3; \dots;\ j_p$ – are the indexes of the $\left|l_{ij}\right|_k^p$ submatrix's lines and columns.

Let's examine the arbitrary dummy $\left|l_{ij}\right|_k^p$ submatrix (fig. 15). Let's choose and cut out from the $\left|l_{ij}\right|_k^p$ submatrix any line and column and calculate the expression: $D_{(k-1),(p-1)} = \sum_{i_1}^{i_{(k-1)}} \sum_{j_1}^{j_{(p-1)}} c_{ij} = 0$

There may be two cases:

1. On the intersection of the cut both column and line in the dummy submatrix there is the $l_{ij} = 0$ element.
2. On the intersection of the cut both column and line in the dummy submatrix there is the $l_{ij} = 1$ element.

In the first case (fig. 16) among all the residuary columns of the $\left|l_{ij}\right|_k^p$ submatrix we may find a $j^{**}$ column of such kind that the number of units in it will remain the same. In some other columns it will be less per unit.

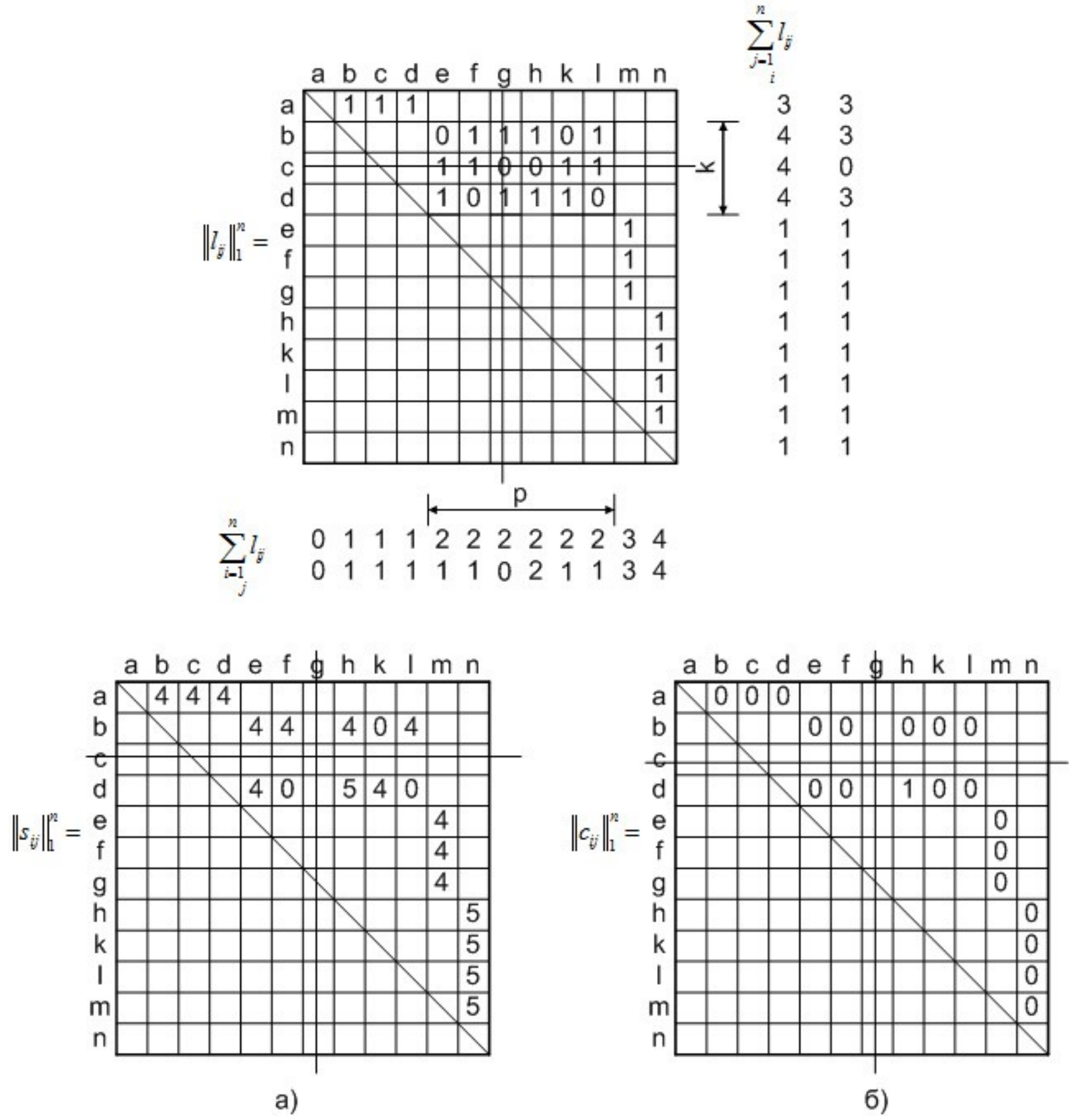

Fig. 16.

Likewise we may find a $i^{**}$ line in which the number of units will remain the same, while in some other lines it will change per unit. Then on the intersection of the $i^{**}$ line and the $j^{**}$ column in the dummy submatrix there will be the $l_{i^{**}j^{**}} = 1$ element, for which: $\max_{ij} s_{ij}^{**} = K_h + P_h$

Among other $l_{ij} = 1$ elements of the dummy submatrix there may be such elements, for which: $\min_{ij} s_{ij} \leq K_h + P_h - 2$. Then in such dummy submatrix without one column and without one line may exist the $c_{i^{**}j^{**}} \neq 0$ value, and therefore, we'll have $D_{(k-1),(p-1)} > 0$.

In the second case (fig. 17) among all the residuary submatrix's columns we will find for sure at least one such $j^{**}$ column, in which the number of units will remain the same, and at least one such column in which the number of units will be less per unit. In exactly the same way, among all the residuary submatrix's lines we will find at least one such $i^{**}$ line, in which the number of units will remain the same, and at least one such line, in which the number of units will be less per unit. Then the dummy submatrix will correspond to at least one element: $\max_{ij} s_{ij}^{**} = K_h + P_h$ and at least one element: $\min_{ij} s_{ij} \leq K_h + P_h - 2$.

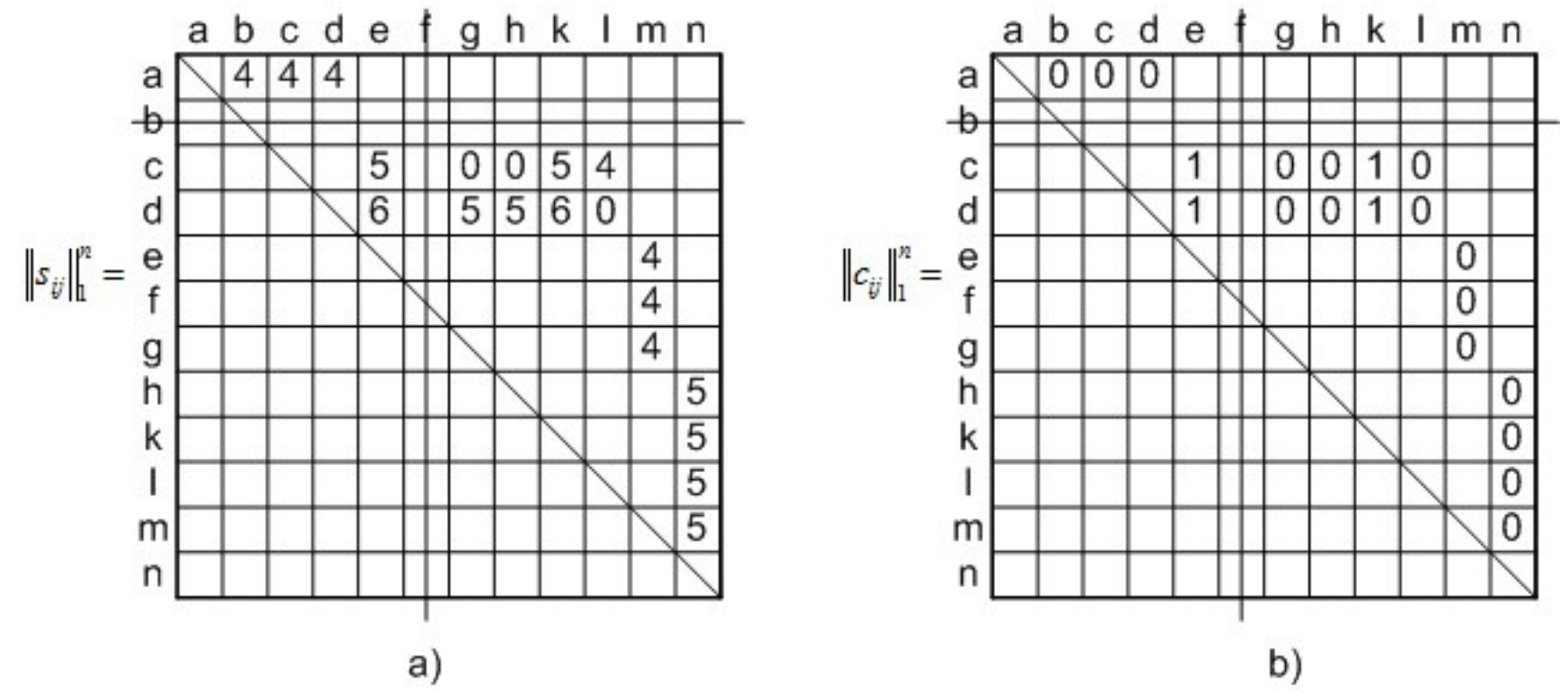

Fig. 17.

So, the dummy submatrix will always correspond to some number of $c_{ij} > 0$ values. In other words, for the dummy submatrix, when we are cutting out of it such lines and columns, that on their intersection the element $l_{ij} = 1$ is located, $D_{(k-1),(p-1)} > 0$ must be.

For example, one of the least dummy submatrixes is the $|l_{ij}|_k^p$ submatrix $(k = 3; p = 3)$, which has zeros along the main diagonal. If we cut both the middle line and the middle column out of this submatrix, then the residuary elements give us all $c_{ij} = 0$, but it is enough to cut the third (first) line and the first (third) column as we get one element $c_{ij} = 2$.

Thus for such submatrix we have $D_{(k-1),(p-1)} > 0$.

So, the dummy submatrix cannot correspond to some $H$ graph's $v_h$ vertex. Therefore, every $l_{ij} = 1$ element belongs to one and only one $H$ graph's $v_h$ vertex if and only if at cutting any such line and any such column out of the $\|l_{ij}\|_1^n$ matrix that in their intersection is located the $l_{ij} = 1$ element, we always have for all the new $\|l_{ij}\|_1^n$ matrix's $l_{ij} = 1$ elements all $c_{ij} = 0$. This is the sufficient condition for every $l_{ij} = 1$ to belong to the quite a definite $v_h$ vertex.

But the calculating of the $c_{ij}$ values, while cutting out of the matrix at the same time of some or other lines and columns, means the calculating of the $c_{ij}$ values for the $\|l_{ij}\|_1^n$ matrix's minors with $(n - 1)$ degree.

After analyzing the two cases of cutting lines and columns out of the dummy submatrix, we can get a final conclusion: it is possible to establish the univocal correspondence for every $l_{ij} = 1$ element of the $L$ matrix, which is considered as the $R$ matrix, to some or other the $H$ graph's vertex, if and only if, the $L$ matrix generates the $\|c_{ij}\|_1^n = [0]$ matrix and the $\|l_{ij}\|_1^{n-1}$ minors for every $l_{ij} = 1$ generate the $\|c_{ij}\|_1^{n-1} = [0]$ matrixes.

Thus an adjacency $L$ matrix (the direct path's matrix) of a $Q = \{q_i\}$ set appears to be the $G$ graph's vertex adjacency $E$ matrix and at the same time it appears to be the $H$ graph's edge adjacency $R$ matrix if and only if:

1. A $L = \|l_{ij}\|_1^n$ generates $C_n = \|c_{ij}\|_1^n = [0]$
2. A minor $|l_{ij}|_1^{n-1}$ of every $l_{ij} = 1$ element generates $C_{n-1} = \|c_{ij}\|_1^{n-1} = [0]$

A theorem is proved.

Let's take up an example in fig. 18. A $L$ matrix appears to be the $R$ matrix because it satisfies to the conditions of theorem 1. A $\|c_{ij}\|_1^n = [0]$ matrix is produced in fig. 19. In this matrix the $c_{ij} = 0$ elements, which correspond to the $l_{ij} = 1$ elements, which enter the same $H(V, Q)$ graph's $v_h$ vertexes, are combined in the rectangular $(k \times p)_h$ submatrixes, which are given the sequence numbers (an algorithm of the numbering will be described later).

$\|l_{ij}\|_1^n = \|\bar{r}_{ij}^{(q)}\| =$

| | a | b | k | d | l | q | m | n | c | f | g | o | p | e | h |
|---|---|---|---|---|---|---|---|---|---|---|---|---|---|---|---|
| a | | 1 | 1 | 1 | 1 | | | | | | | | | | |
| b | | | | | | 1 | 1 | 1 | | | | | | | |
| k | | | | | | | | | 1 | | | | | | |
| d | | | | | | | | | | 1 | 1 | | | | |
| l | | | | | | | | | | | | | | | 1 |
| q | | | | | | | | | 1 | | | | | | |
| m | | | | | | | | | | | | | | 1 | |
| n | | | | | | | | | | | | | | | 1 |
| c | | | | | | | | | | 1 | 1 | | | | |
| f | | | | | | | | | | | | 1 | 1 | | |
| g | | | | | | | | | | | | | | 1 | |
| o | | | | | | | | | | | | | | 1 | |
| p | | | | | | | | | | | | | | | 1 |
| e | | | | | | | | | | | | | | | 1 |
| h | | | | | | | | | | | | | | | |

Fig. 18. An ***L*** matrix, which satisfies to the conditions of theorem 1

Fig. 19 has the following designations:

$N_{hn}$ – A $H$ graph's $v_h$ vertex number, from which the $q_i$ edge comes out;

$N_{hk}$ – A $H$ graph's $v_h$ vertex number, in which the $q_i$ edge comes in;

An integration of the $l_{ij} = 1$ elements into the submatrixes and their numeration allows arranging the $F$ matrix (Fig 20). The row's indexes are the submatrix's numbers in the $\|l_{ij}\|_1^n$ matrix.

**Definition 2.3.** A direct path's $\|l_{ij}\|_1^{n_q}$ matrix, which satisfies to conditions of the theorem 1 (a quasicanonical adjacency matrix's theorem) is called the *quasicanonical adjacency matrix*. This matrix at the same time is the $\|e_{ij}\|_1^{n_q}$ matrix – the $G_q(Q_q, \Gamma_q)$ graph's vertex adjacency matrix (fig. 21a) and the $H_q(V_q, Q_q)$ graph's edge adjacency matrix (fig 21b), and thus has the dual nature.

**Remark.** We put index $q$ in cases, when we talk about quasicanonical objects.

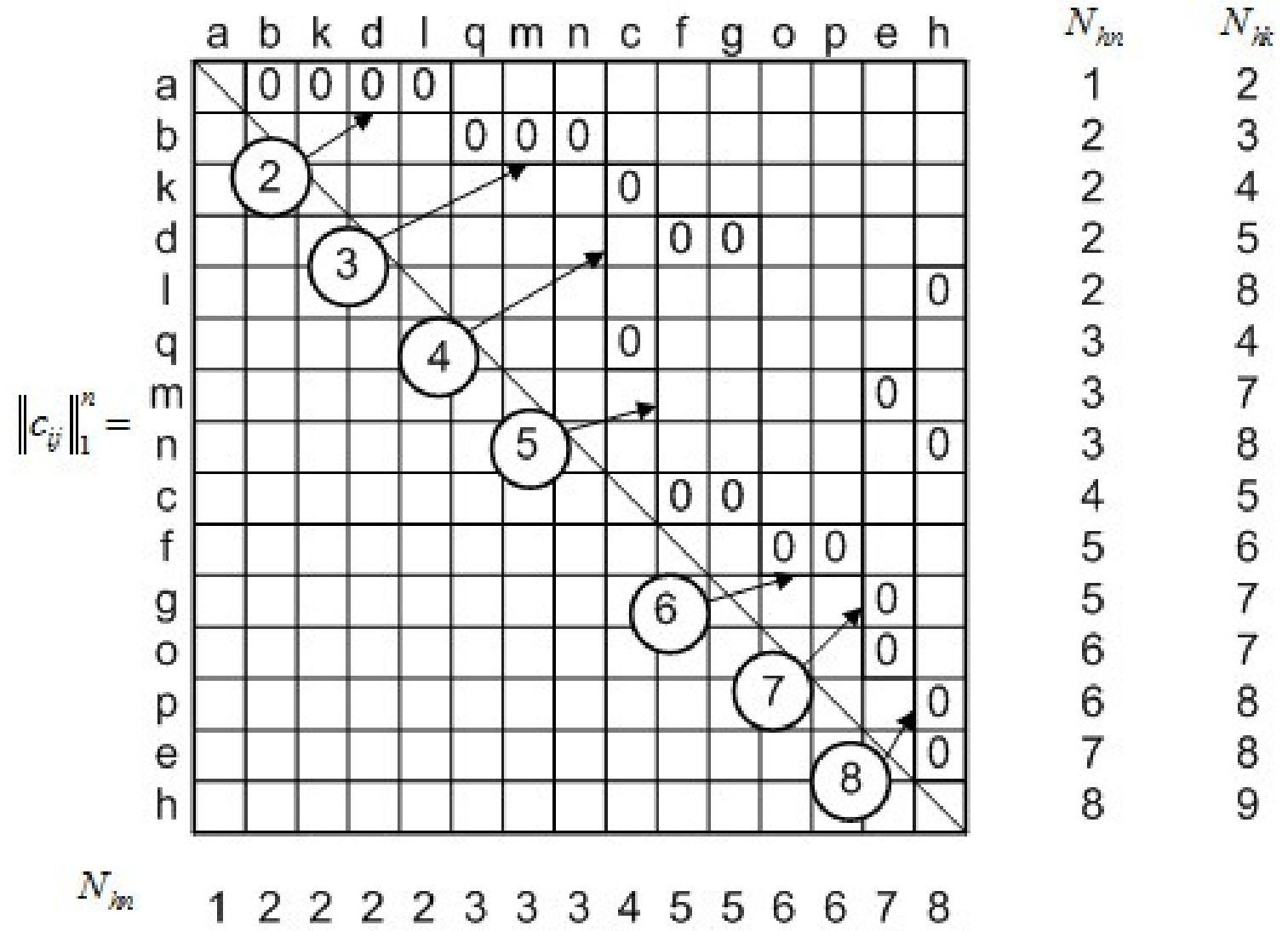

Fig. 19. The $\|\boldsymbol{c_{ij}}\|_{\mathbf{1}}^{\boldsymbol{n}} = [\mathbf{0}]$ matrix

$\|f_{nd}\|_1^9 =$

| | 1 | 2 | 3 | 4 | 5 | 6 | 7 | 8 | 9 |
|---|---|---|---|---|---|---|---|---|---|
| 1 | | a | | | | | | | |
| 2 | | | b | k | d | | | l | |
| 3 | | | | q | | | m | n | |
| 4 | | | | | c | | | | |
| 5 | | | | | | f | g | | |
| 6 | | | | | | | o | p | |
| 7 | | | | | | | | e | |
| 8 | | | | | | | | | h |
| 9 | | | | | | | | | |

Fig. 20. An $\boldsymbol{F}$ matrix

Let's call the $G_q$ graph as the quasicanonical vertex graph, and the $H_q$ graph as the quasicanonical edge graph. A $G_q$ graph in this case appears to be adjacent for the $H_q$ graph's edges, or the conjugate graph for the $H_q$ graph. The graph's cyclomatic numbers always satisfy to the condition: $\nu(G_q) \geq \nu(H_q)$.

It reflects the possibility of the presence in the $H_q$ graph the complicated vertexes. Later such circumstances will be examined in full. Besides, this condition mirrors the degeneracy of the duality (quasiduality) of the quasicanonical adjacency matrix.

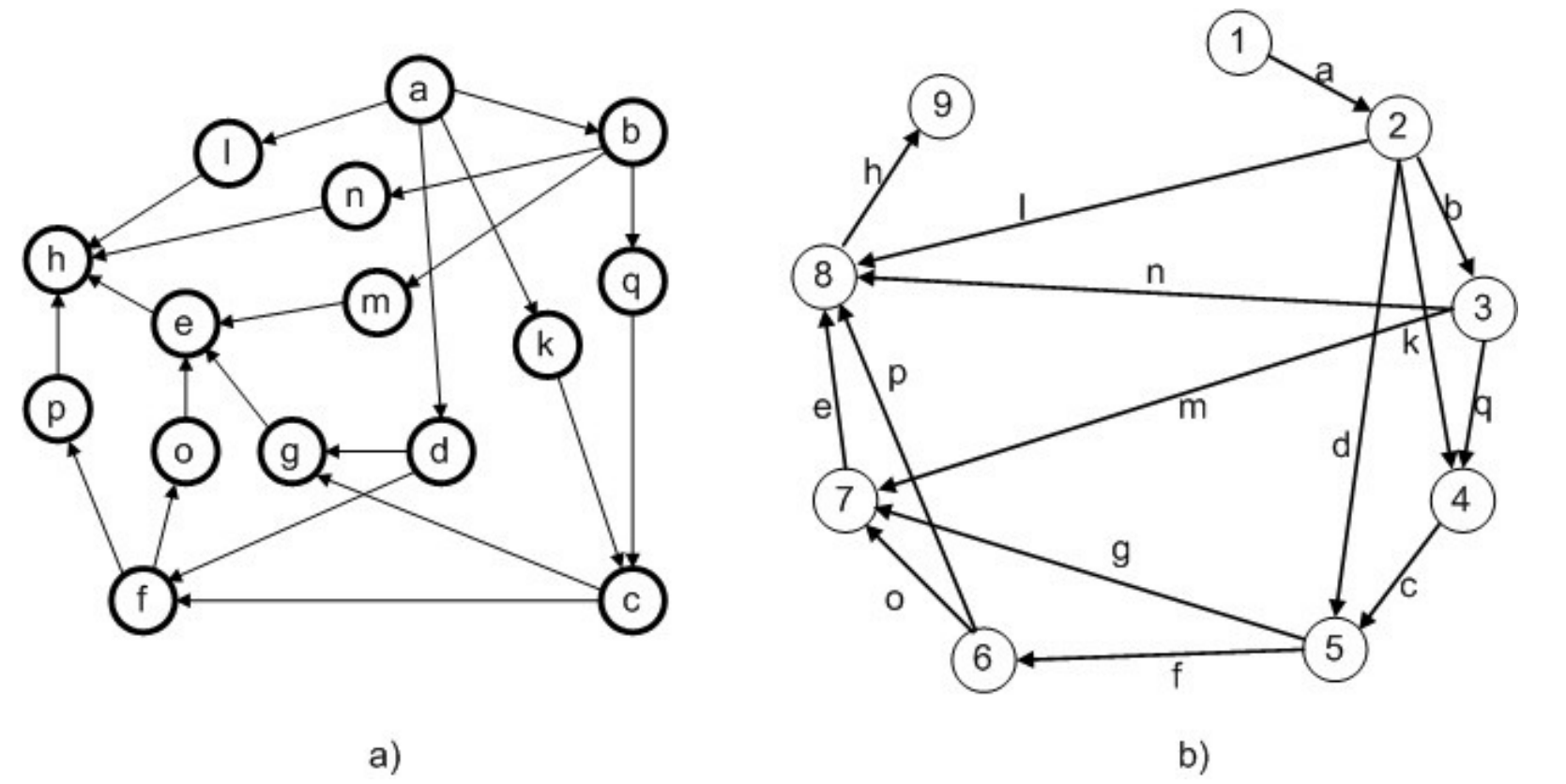

Fig. 21. The equivalent both the vertex $\boldsymbol{G_q}(\boldsymbol{Q_q}, \boldsymbol{\Gamma_q})$ and the edge $\boldsymbol{H_q}(\boldsymbol{V_q}, \boldsymbol{Q_q})$ graphs

In our example: $\nu(G_q) = 8$, and $\nu(H_q) = 7$

In a particular case, when $\nu(G_q) = \nu(H_q)$, the $L_q = R_q$ matrix is called canonical or normal.

So, the conditions of the existence of the quasicanonical adjacency matrix were examined in theorem 1, though in practice you can meet such matrixes only by chance. Thus it becomes important to find a method of transforming every direct path's matrix, which does not satisfy to conditions of theorem 1 to the needed form. The transformation of the $L$ matrix must be of such a form so that the binary relation's system was invariable. That is: the transformation must be conservative to the binary relation's system, which is adjusted on the $Q = \{q_i\}$ set.

# 3 A transformation of the direct path's matrix to the quasicanonical form

**Definitions:**

## 3.1 A conservative transformation of the binary relation

By the conservative transformation of the binary relation between the two $q_i$ and $q_j$ elements we'll denote such a transformation, which will allow either to insert the additional elements into the $Q = \{q_i\}$ set or to exclude them without changing the relation between the $(q_i; q_j)$ elements. Such a transformation may be based on the transitivity property of the binary relation. For example, the initial pair is defined as $(q_i; q_j) \in Q$. Let the elements be connected by the relation: $q_i < q_j$. Let's accept two conditions: both $q_i < q_z$ and $q_z < q_j$, and transform the initial expression. We'll find that $q_i < q_z < q_j$. It is obvious, that two relations both $q_i < q_j$ and $q_i < q_z < q_j$ are equivalent according to the initial pair of the elements. Therefore, the inserting of the $q_z$ element into the $q_i < q_j$ relation – is the conservative operation regarding to this relation in the initial pair.

### 3.2 A Δn-transformation of the $L$ matrix

Let us settle that by the $\Delta n$ - transformation of the $L$ matrix we will comprehend the addition of one row (both line and column) to the $L$ matrix at the condition of replacement the $q_x < q_y$ relation with the pair of binary both $q_x < q_{n+1}$ and $q_{n+1} < q_y$ relations. It is evident, that the $\Delta n$ - transformation is the conservative operation regarding the binary relation in the initial $(q_i; q_j)$ pair and does not break such a structural similarity criterion as the binary relation's system.

### *Theorem 2 "On a quasinormalization of the $L$ matrix's binary relations"*

Any direct path's $\|l_{ij}\|_1^n$ matrix can be transformed to the quasicanonical (quasinormal) $\|l_{ij}\|_1^{n+s_q}$ form by means of applying the $\Delta n$ - transformation to such $s_q$ ($s_q \leq n^2 - 1$) elements of the $\|l_{ij}\|_1^n$ matrix, which do not satisfy to the conditions of theorem 1.

**Proof:**

An arbitrary $L$ matrix is given. It does not satisfy to the conditions of theorem 1 (fig.22). Let's mark such elements, which do not satisfy to the conditions of theorem 1 in the $L$ matrix, in other words, which generate ($c_{ij} \neq 0$), with the help of circles.

Let us accomplish the $\Delta n$ - transformation to the every element that does not satisfy to the conditions of theorem 1 (fig. 23a):

1. We expel the marked $l_{xy} = 1$ element, for example, $l_{ac} = 1$, from the $\|l_{ij}\|_1^n$ matrix (fig.22a).

2. Then we'll add one row, which is both the line and the column, $l = j = k$ to the $\|l_{ij}\|_1^n$ matrix (fig. 23a).

3. Then we'll add the two $l_{x,(n+1)} = 1$ and $l_{(n+1),y} = 1$ elements to the $\Gamma: Q \to Q$ set (fig. 23a: $l_{ak} = 1$ and $l_{kc} = 1$).

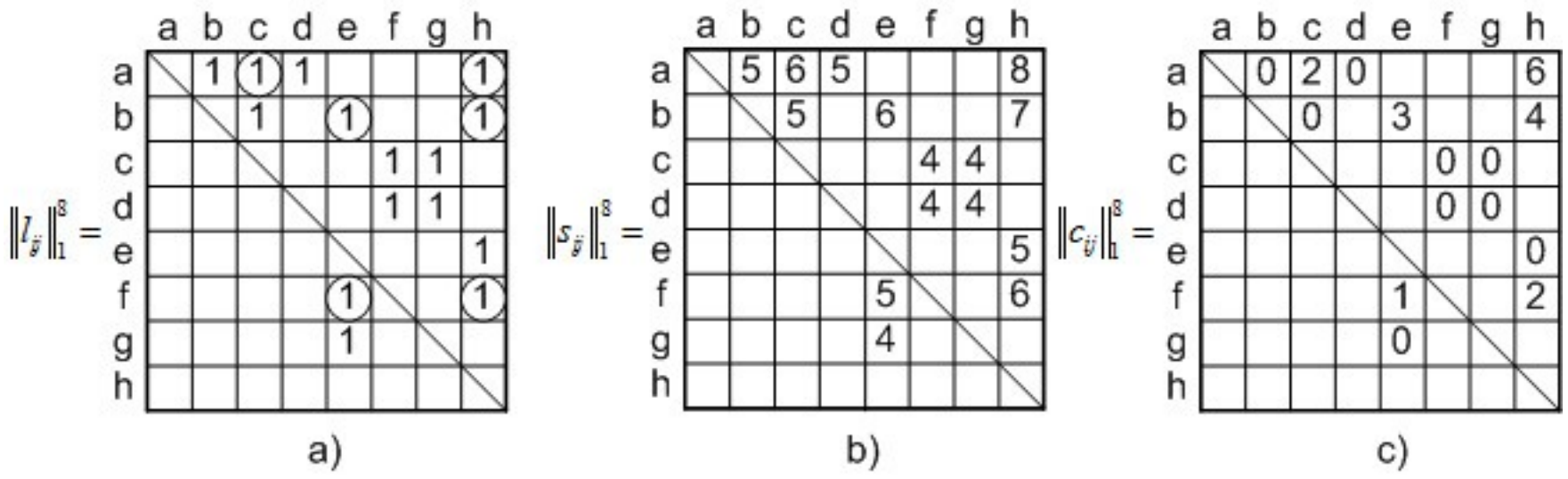

Fig. 22. First calculation of $\|\boldsymbol{c_{ij}}\|_1^n$ matrix

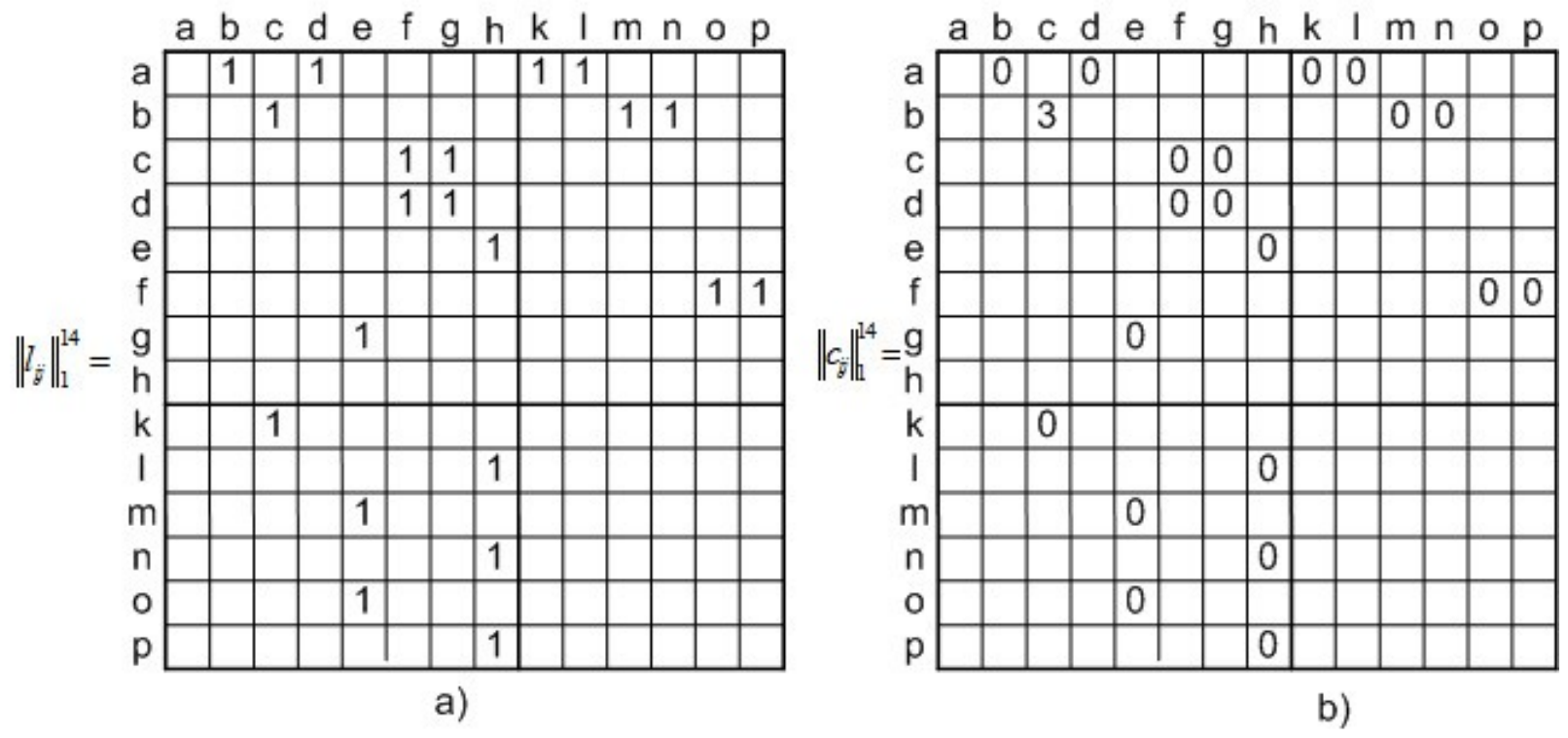

Fig. 23. An application of the $\boldsymbol{\Delta n}$- transformation to every $\boldsymbol{L}$ matrix's elements, which do not satisfy to the conditions of theorem 1

As a result, instead of each $l_{xy} = 1$ element, which does not satisfy to theorem 1 conditions and the $q_x < q_y$ relation, we'll get the two elements: $l_{x,(n+1)} = 1$ and $l_{(n+1),y} = 1$, and the two relations: $q_x < q_{n+1}$ and $q_{n+1} < q_y$, which give us the $q_x < q_{n+1} < q_y$ relation, equivalent to the initial $q_x < q_y$ relation (fig.23a). At the same time the new $l_{x,(n+1)} = 1$ and $l_{(n+1),y} = 1$ elements satisfy to theorem 1 conditions, that is $c_{x,(n+1)} = 0$ and $c_{(n+1),y} = 0$ are corresponding to them. After all of the $l_{ij} = 1$ elements, which do not satisfy to the conditions of theorem 1 and which were discovered during the first control, are subjected to the $\Delta n$ - transformation, we'll check-up a matrix for satisfying to the conditions of theorem 1 (fig. 23b) once more. Again we'll accomplish, if necessary, the $\Delta n$ - transformation. We'll do this job till all the $l_{ij} = 1$ elements, which do not satisfy to the conditions of theorem 1, are not revealed. As a result, a matrix will get the quasicanonical form (fig. 24).

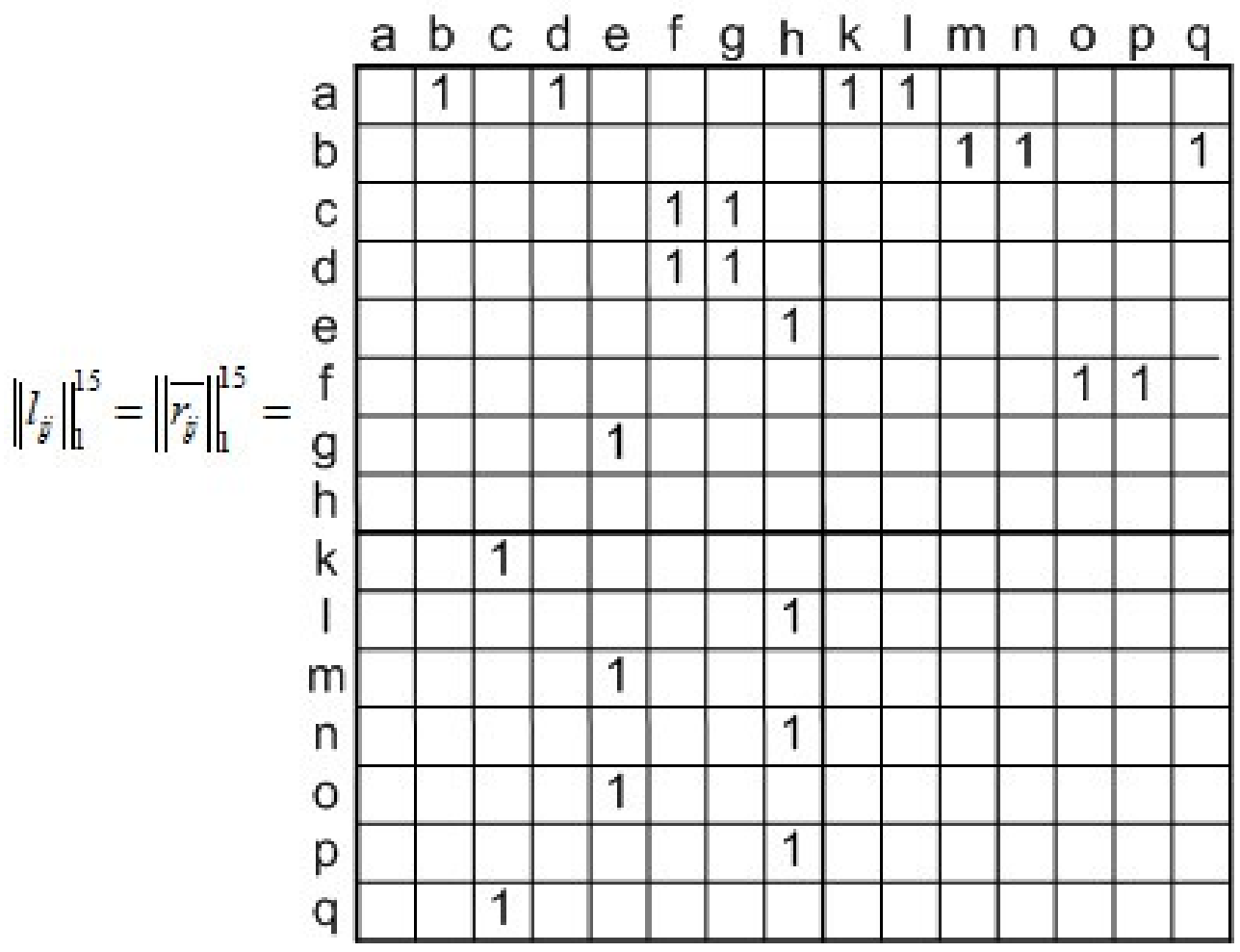

Fig. 24. A quasicanonical $\left\|\boldsymbol{l_{ij}}\right\|_{\boldsymbol{1}}^{\boldsymbol{15}}$ matrix

A matrix which was subjected to the $\Delta n$ - transformation at least once has two groups of rows: both initial and additional. In every additional row there is only one element: $l_{x,(n+1)} = 1$ and $l_{(n+1),y} = 1$. That's why none of them can generate $c_{ij} \neq 0$.

A number of the elements in the initial rows after the $\Delta n$ - transformation of one element is decreasing per unit. Thus, a number of the $l_{ij} = 1$ elements, which correspond to $c_{ij} \neq 0$ in either the $\|c_{ij}\|_1^n$ matrix or in the $C_n = \|c_{ij}\|_1^{n+1}$ matrix, tend to zero as the $\Delta n$ - transformation is applied. The biggest number of the elements, which may be subjected to the $\Delta n$ - transformation, can not be more then the biggest $L$ matrix's number of the $l_{ij} = 1$ elements, which does not satisfy to the conditions of theorem 1. This number appears to be $(n^2 - 1)$ equal, since at the number of the $l_{ij} = 1$ elements equal to $n^2$ the matrix $\|l_{ij}\|_1^n$ satisfies to the conditions of theorem 1.

So, the process of the consistent $\Delta n$ - transformation of the $L$ matrix appears to be converging.

**Remark.** While proving theorem 1, we do not apply the restrictions to the fact of either presence or absence of the loops in the $G$ graph. The $l_{ij} = 1$ elements, which correspond to the loops in the $G$ graph, may be subjected to the $\Delta n$ - transformation as well as any other $l_{ij} = 1$ elements. So, a loop, consisting of one edge and one vertex, is transformed into a contour, which consists of two edges and two vertexes. In real practice such a contour represents the alternation of the reiterations of some process' elements with the signals, which allow these reiterations (a finite number of times). Thus, the $\Delta n$ - transformation of the $l_{ii} = 1$ elements also correspond to the transitivity property between the elements of the real process.

Since the $\Delta n$ - transformation may be applied to every $l_{ij} = 1$ element without the disturbance of the initial system of the binary relations, so every finite $L$ matrix can be transformed to the quasicanonical (quasinormal) form with the help of the $\Delta n$ - transformation to those, but not more then $(n^2 - 1)$, elements, which do not satisfy to the conditions of theorem 1.

The theorem is proved.

Let's call the process of such matrix's transformation as quasinormalization.

A transformed matrix's order will be: $n_q = n + s_q \leq n + n^2 - 1 = n(n + 1) - 1$

In the investigated example the process of the $\Delta n$ - transformation converges in two steps (fig. 24). A number of the transformed matrix's elements will be:

$\sum_{i=1}^{n_q} \sum_{j=1}^{n_q} l_{ij}^{(q)} = \sum_{i=1}^{n} \sum_{j=1}^{n} l_{ij} + s_q$

Since $s_q \leq n^2 - 1$, then:

$\sum_{i=1}^{n_q} \sum_{j=1}^{n_q} l_{ij}^{(q)} = \sum_{i=1}^{n} \sum_{j=1}^{n} l_{ij} + n^2 - 1$

Let's return to the investigated example of reforming the $L$ matrix to the quasicanonical form and give some explanations. The initial $G$ graph, corresponding to the $\|l_{ij}\|_1^n$ matrix, which is presented in fig. 22a, is represented in fig. 25a.

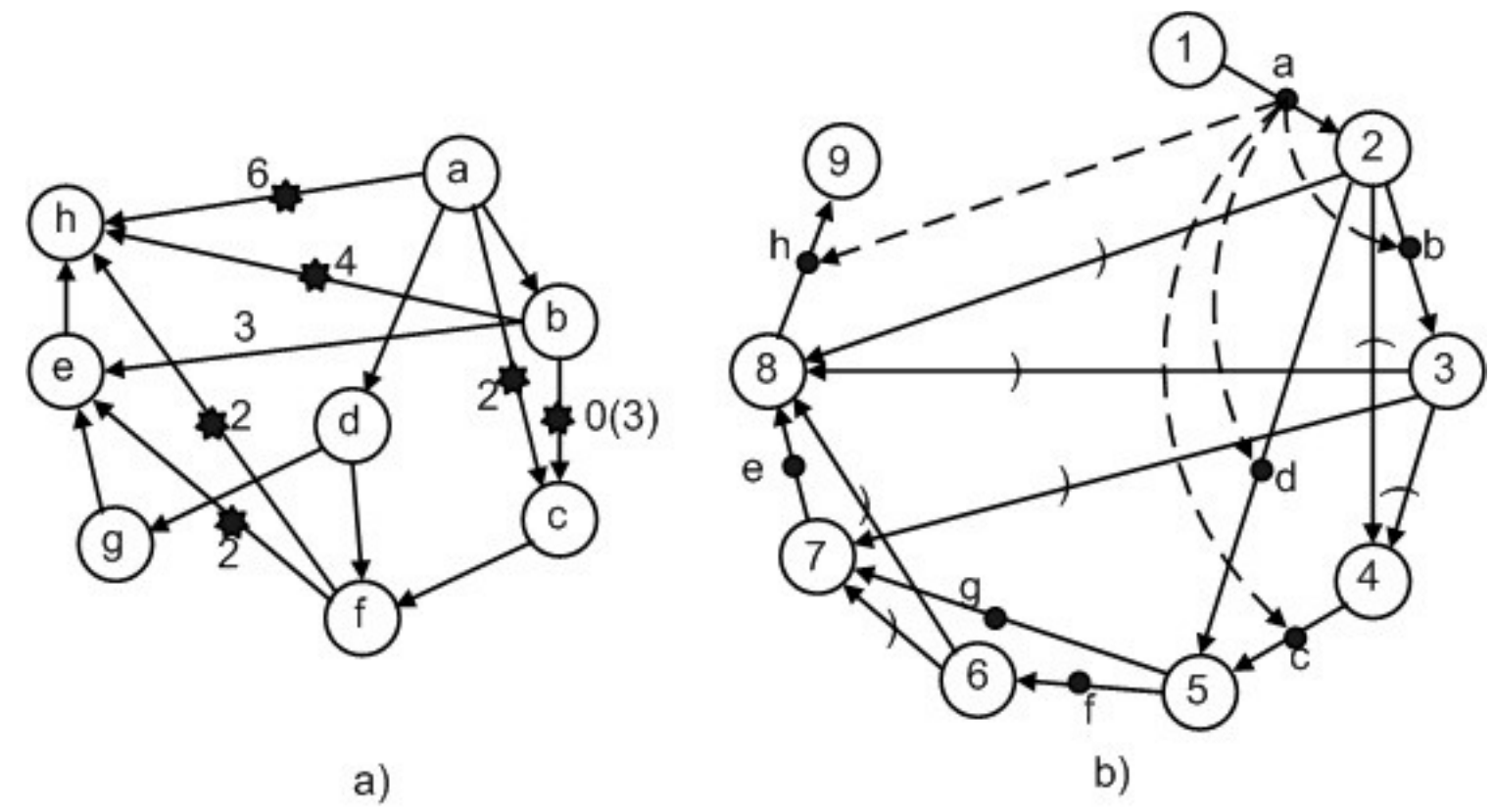

Fig. 25 a) Initial $\boldsymbol{G}(\boldsymbol{Q},\boldsymbol{\Gamma})$ graph; b) A $\boldsymbol{H_q}(\boldsymbol{V_q},\overline{\boldsymbol{Q_q}})$ graph and a fragment of the initial $\boldsymbol{G}(\boldsymbol{Q},\boldsymbol{\Gamma})$ graph

The $l_{ij} = 1$ elements, which need the $\Delta n$ – transformation, are marked by the circles on the initial $L$ matrix (fig. 22a). It is the result of the first calculation of the $\|c_{ij}\|_1^n$ matrix. The initial $G$ graph's (fig. 25a) edges, which need the $\Delta n$ – transformation, are marked by the stars with the numbers equal to $c_{ij} \neq 0$. The edge $(b, c)$ is marked by a star with zero and number 3 in the brackets, because we have $c_{bc} = 3$ after the first step of the $\Delta n$ – transformation.

The $\Delta n$ – transformation has the very simple graphic interpretation on the initial graph. The edges, for which $c_{ij} \neq 0$, are divided into two consecutive ones, and the new additional vertexes are inserted into the gaps. All the edges, which have $c_{ij} \neq 0$, are divided. After this operation for all the edges, which are adjacent to the divided edges, the $c_{ij}$ value is calculated once more, and again the edges, which have $c_{ij} \neq 0$, are divided in two consecutive ones. The process of the $\Delta n$ - transformation continues up to the moment, when the $c_{ij}$ values will be equal to zero for all the edges. It is evident that the $\Delta n$ – transformation may be started with any $l_{ij} = 1$ element, which has the $c_{ij} \neq 0$ value, and thus, from any of the corresponding edges in the initial graph. The solution would not change because of that.

After the two steps of the $\Delta n$ – transformation the $L$ matrix is brought to the quasicanonical $\|r_{ij}^{(q)}\|_1^{15}$ form (fig. 24). A matrix $\|r_{ij}^{(q)}\|_1^{15}$ is presented in the already ordered form in fig 18.

A quasicanonical matrix permits to form the $F_q$ matrix – the $H_q$ graph's vertex adjacency matrix (fig. 20) and construct two quasicanonical graphs: $G_q$

(fig. 21a) and $H_q$ (fig. 21b). A $H_q$ graph contains all the necessary information about the initial $G_q$ graph (fig. 25b).

Let's mark the centers of that $H_q$ graph's edges, which are part of the initial $Q$ set. Having accepted such points for the vertexes and connecting them by edges in compliance with the initial $L$ matrix, we'll get the initial $G$ graph. The initial $G$ graph's fragment is displayed with the help of the dotted arrows in fig. 25b.

So, the initial graph's edges appear to be the adjacent edges of the $H_q$ graph's edges, but this contiguity is of two kinds. In some cases this contiguity is direct, in others – through the additional edge, which was brought in by the $\Delta n$ -transformation. Let's call such contiguity as *transit contiguity*. Then the $G$ graph appears to be the partially transit contiguity graph to the $H_q$ graph's edges. In a particular case, when all the $l_{ij} = 1$ elements of the $L$ matrix undergo the $\Delta n$ -transformation, the initial $G$ graph appears to be a completely transit contiguity graph of the $H_q$ graph's edges.

As it was mentioned above, a quasicanonical $R_q$ matrix generally does not allow having the $H_q$ graph with the same cyclomatic number, which the initial $G$ graph has. This problem is of particular interest and it would be examined later.

**Remark.** A transformation of the $L$ matrix into the $R_q$ matrix is accompanied by bringing the new elements into the initial set. In the real process such elements play the role of the fictitious jobs (the additional elements correspond to the fictitious jobs in an ordinary net models, and in a conjugated canonical net model they obtain the meaning of the effective jobs) [6]. Thus, the main set, if we are talking about an ordinary net model of a real process, includes the initial set, which corresponds to either the effective jobs or the operations, and its supplement, which corresponds to either the fictitious jobs or enabling signals. Such circumstances do not destroy the structural similarity of the real process and its model:

1. The number of the elements of the initial set does not change.
2. The numbers of the elements of the main set of graphs of both forms are equal.
3. An addition to the initial set (the fictitious edges or vertexes) has quite a real physical interpretation in the actual process as a set of the enabling signals.

# 4 The canonical vertex and edge graphs

Let's examine the case of the strict duality of both vertex and edge graphs, when the demand of the cyclomatic number's equality applies together with other structural similarity criteria. Let us also make some remarks according the cyclomatic number. Either a cyclomatic number or a maximal number of the connected $G(Q, \Gamma)$ graph's independent cycles may be determined with the help of the adjacency matrix, if each undirected edge is replaced by two directed ones [2].

A number of the $G(Q, \Gamma)$ graph's edges are equal: $\nu(\Gamma) = \sum_{i=1}^{n} \sum_{j=1}^{n} l_{ij}$

And, consequently, the $G(Q, \Gamma)$ graph's cyclomatic number will be:

$\nu(Q) = \sum_{i=1}^{n} \sum_{j=1}^{n} l_{ij} - n + p$

Where:

$n$ – is an $L$ matrix's order and, at the same time, a number of the $G$ graph's vertexes;

$p$ – is the number of the $G$ graph's components of the connectivity.

On the other hand, the sum – $\sum_{i=1}^{n} \sum_{j=1}^{n} l_{ij}$ – is a number of the $G$ graph's edges and, at the same time, the number of all ordered couples of elements in the binary relation's $L = Q \times Q$ system, which is adjusted on the $Q = \{q_i\}$ set.

So, the cyclomatic number is a quantitative estimation of the binary relation's $L = Q \times Q$ system. In other words, $\nu(Q)$ is a cyclomatic number of $p$-connected $Q = \{q_i\}$ set with the prescribed binary relation's $L = Q \times Q$ system.

However, if a graph represents the given binary relation's system, then the cyclomatic number must not depend on the graph's appearance. Thus, we have a theorem on the equality of the cyclomatic numbers of the graph $G$ and the graph $G_q$.

## *Theorem 3 "On the equality of both vertex and edge graph's cyclomatic numbers"*

A cyclomatic $\nu(Q)$ number of both the initial vertex $G$ graph and the quasicanonical $G_q$ graph, which is received by the way of the initial graph's quasinormalization, are equal.

**Proof**

Indeed, the equal number of both vertexes and edges is added to the initial vertex $G$ graph by every step of the $\Delta n$- transformation. As a result, the $G$ graph's cyclomatic number during its transformation to the $G_q$ graph remains constant.

Theorem is proved.

Let's bring in some definitions.

**Definition 4.1.** By the $H$ graph's simple $v_h$ vertex we'll denote such vertex, which corresponds to the $|r_{ij}|_k^p$ submatrix of the $R$ matrix at both $k \geq 1$ and $p = 1$ or at both $k = 1$ and $p \geq 1$.

**Definition 4.2.** By the $H$ graph's elementary $v_h$ vertex we'll denote such a vertex which corresponds to the $|r_{ij}|_k^p$ submatrix of the $R$ matrix at both $k = 1$ and $p = 1$.

**Definition 4.3.** By the $H$ graph's complicated $v_h$ vertex we'll denote such $v_h$ vertex which correspond to the $|r_{ij}|_k^p$ submatrix of the $R$ matrix at both $k \geq 2$ and $p \geq 2$.

## *Theorem 4 "On the canonical adjacency matrix"*

Let the $G$ initial graph be specified as the $\|e_{ij}\|_1^n$ matrix – an adjacency matrix of vertexes, which corresponds to the quasicanonical $\|r_{ij}\|_1^{n+s_q}$ matrix – an adjacency matrix of edges of the connected edge $H_q$ graph. In order that the $H_q$ graph's cyclomatic $\nu(H_q)$ number might be equal to the initial $G$ graph's cyclomatic $\nu(G)$ number, it is necessary and sufficient for all the $H_q$ graph's vertexes to be simple, or, for every $r_{xy} = 1$ such condition must be fulfilled:

**Condition 1**

$$\left.\begin{array}{l} \text{If } \sum_{\substack{i=1 \\ j=y}}^{n+s_q} r_{ij} \geq 1, \text{then } \sum_{\substack{j=1 \\ i=x}}^{n+s_q} r_{ij} = 1 \\ \text{If } \sum_{\substack{j=1 \\ i=x}}^{n+s_q} r_{ij} \geq 1, \text{then } \sum_{\substack{i=1 \\ j=y}}^{n+s_q} r_{ij} = 1 \end{array}\right\}$$

**Proof**

Every independent cycle may be examined as the closed consecution of the ordered couples of the $Q$ set's elements (irrespective of the orientation of these pairs). Owing to the fact that the same binary relation's system, being fixed on the $Q = \{q_i\}$ set, may be expressed by the $\|e_{ij}\|_1^n$ matrix – an adjacency matrix of $G$ graph's vertexes and by the $\|r_{ij}\|_1^{n+s_q}$ matrix – an adjacency matrix of the $H_q$ graph's edges, the exclusion of either one or other element from the $L = Q \times Q$ set must equally change the cyclomatic number of both the $G$ graph and the $H_q$ graph. This condition may be represented as:

**Condition 2**

$$\frac{\Delta\nu(G)}{\Delta\left(\sum_{i=1}^{n}\sum_{j=1}^{n} l_{ij}\right)} = \frac{\Delta\nu(H_q)}{\Delta\left(\sum_{i=1}^{n}\sum_{j=1}^{n} r_{ij}\right)}$$

The theorem's proof leads to the proving of:

1. Condition (2) is right in that and only that case, when all the $H_q$ graph's vertexes are simple, that is – condition (1) is right.

2. The cyclomatic numbers of both the $H_q$ graph and the $G$ graph, constructed on the base of one binary relation's system, are equal if and only if condition (1) is also right.

Let us prove the theorem for the case of the connected graphs.

Let us denote:

$$\frac{\Delta\nu(G)}{\Delta\left(\sum_{i=1}^{n}\sum_{j=1}^{n} l_{ij}\right)} = \delta_\nu(G);$$

$$\frac{\Delta\nu(H_q)}{\Delta\left(\sum_{i=1}^{n}\sum_{j=1}^{n} r_{ij}\right)} = \delta_\nu(H_q);$$

For the determination of both $\delta_\nu(G)$ and $\delta_\nu(H_q)$ values we'll examine how the exclusion of either one or other $l_{xy} = 1$ element from the $L = Q \times Q$ set influences both the cyclomatic number of both the $G$ graph and the $H_q$ graph.

These variants take place for the $G$ graph:

- The $G$ graph appears to be a tree, then $\nu(G) = 0$;

- The $G$ graph appears not to be a tree, then $\nu(G) \geq 1$, but the excluded $l_{xy} = 1$ element is congruent with the divided edge;
- The $G$ graph appears not to be a tree, then $\nu(G) \geq 1$, but the excluded $l_{xy} = 1$ element is not congruent with the divided edge.

In first two cases the exclusion of the $l_{xy} = 1$ element divides the $G$ graph into two disconnected components, so we do not examine such variants. In the third case the $G$ graph remains to be connected, but at the exclusion of the $l_{xy} = 1$ element one of the independent cycles becomes breaked. Therefore, $\Delta\nu(G) = -1$ and $\delta_\nu(G) = 1$. So, on the condition of the preservation of the graph's connectedness, the term is always right: $\delta_\nu(G) = 1$. So, for the fairness of condition (2), that is: $\delta_\nu(G) = \delta_\nu(H_q)$, the equality of $\delta_\nu(H_q) = 1$ for the $H_q$ graph is also necessary at the excluding from the binary relation of such elements, which exclusion leads to $\delta_\nu(G) = 1$.

All the $H_q$ graph's vertexes may be divided into two classes: the simple vertexes and the complicated vertexes (the elementary vertexes are included in a number of the simple vertexes). The $l_{xy} = 1$ element, which is corresponding to the $q_x < q_y$ relation, and which exclusion from the $G$ graph, will bring $\Delta\nu(G) = -1$; may enter the simple vertex, as well as the complicated vertex of the $H_q$ graph. An initial $q_x < q_y$ relation may enter the $\|r_{ij}\|_1^{n+s_q}$ matrix either in the initial $q_x < q_y$ form or in a form of $q_x < q_\alpha < q_y$ relation, if the initial relation was exposed to the $\Delta n$ – transformation.

For the exclusion of the initial relation from the $\|r_{ij}\|_1^{n+s_q}$ matrix it is sufficient to expel the $r_{xy} = 1$ element. If we have the $q_x < q_\alpha < q_y$ relation already transformed, we can exclude either the $r_{x\alpha} = 1$ element or the $r_{\alpha y} = 1$ element. Generally speaking, it is all the same.

Let's examine the case of the $q_x < q_y$ relation, which enters the simple $H_q$ graph's vertex. We have variants:

- The $H_q$ graph appears to be a tree, then $\nu(H_q) = 0$;
- The $H_q$ graph appears not to be a tree, then $\nu(H_q) \geq 1$, but the excluded $r_{xy} = 1$ element is the dividing, in other words, at least one of the two edges either $q_x$ or $q_y$ appears to be the dividing ones;
- The $H_q$ graph appears not to be a tree, then $\nu(H_q) \geq 1$ and the $r_{xy} = 1$ element is not the dividing one, so neither $q_x$ edge nor $q_y$ edge appears to be the dividing one.

In the first two cases the exclusion of the $r_{xy} = 1$ element makes the graph disconnected, that's why these variants are not examined. In the last variant the exclusion of $r_{xy} = 1$ leads to an additional appearance of either one dangling vertex or two dangling vertexes instead of one vertex, with the sum of the in-degrees equal 2, at the same number of the edges. It is equal to the breaking of one of the independent cycles, that is: $\nu(H_q) = -1$ and, so: $\delta_\nu(H_q) = 1$

Thus, if the excluding relation enters the simple vertex, then, by the condition of the preservation of the $H_q$ graph's connectedness, the condition $\delta_v(H_q) = 1$ is always right.

Let the $r_{xy} = 1$ element enter the $H_q$ graph's complicated vertex. There may also take place such variants as in the first case. Just as before we'll examine only the third variant, when the $H_q$ graph appears not to be a tree and neither of $q_x$ or $q_y$ edges are the dividing edges. In this case the exclusion of the $r_{xy} = 1$ element breaks the $\|r_{ij}\|_1^{n+s_q}$ matrix's the quasicanonical quality. Then, for the purpose of bringing the matrix to the quasicanonical form, it is necessary to do the $\Delta n$ – transformation again.

Let us examine the typical cases:

- A $|r_{ij}|_k^p$ submatrix, corresponding to the $v_h$ vertex has $k = 2$ and $p = 2$. After the exclusion of the $r_{xy} = 1$ element it is necessary to do the $\Delta n$-transformation of one $r_{xy} = 1$ element in order to bring the matrix to the quasicanonical form. As a result the extra vertex and the extra edge are generated. Then: $\delta_v(H_q) = 0 \neq 1$
- A $|r_{ij}|_k^p$ submatrix, corresponding to the $v_h$ vertex has $k = 2$ and $p = 3$ or $k = 3$ and $p = 2$. After the exclusion of the $r_{xy} = 1$ element it is necessary to do the $\Delta n$- transformation of the four $r_{xy} = 1$ elements in order to bring the matrix to the quasicanonical form. As a result a number of the $H_q(V_q, Q_q)$ graph's edges will increase by four, and a number of vertexes – by three. Then: $\delta_v(H_q) = {}^1/_4 \neq 1$
- A $|r_{ij}|_k^p$ submatrix, corresponding to the $v_h$ vertex has $k \geq 3$ and $p \geq 3$. After the exclusion of the $r_{xy} = 1$ element it is necessary to do the $\Delta n$- transformation of $(k * p - 1)$ the $r_{ij} = 1$ elements in order to bring the $\|r_{ij}\|_1^{n+s_q}$ matrix to the quasicanonical form. As a result, the $(k * p - 1)$ edges and the $(k + p - 1)$ vertexes will be added to the $H_q(V_q, Q_q)$ graph. A number of the elements in the matrix will increase by $(k * p - 1)$ units.

Then, meaning, that for the connected $H_q$ graph: $\nu(H_q) = n + s_q - \nu(V_q) + 1$, where: $\nu(V_q)$ – is a number of the graph's vertexes, we'll get:

$$\delta_v(H_q) = \frac{k*p-k-p}{k*p-1}$$

It is obvious, that $\delta_v(H_q) \neq 1$ is right at any whatsoever $k \geq 3$ and $p \geq 3$. Indeed, for the purpose of $\delta_v(H_q) = 1$ it is necessary that $k + p = 1$, and it is impossible.

So, if the excluding $q_x < q_y$ relation enters the complicated vertex, then always: $\delta_v(H_q) \neq 1$

So for keeping condition (2) of the theorem 4 or the condition: $\delta_v(G) = \delta_v(H_q) = 1$ it is sufficient for the graph's vertexes to be simple.

But this condition is also a necessary condition. Indeed, if the $q_x < q_y$ relation enters the $H_q$ graph's complicated vertex, then in the $G$ graph it enters the cycle and, so its exclusion always generates $\delta_v(G) = 1$ and couldn't generate $\delta_v(G) = 0$.

Therefore, for providing the fulfillment of the conditions either (2) or ($\delta_v(G) = \delta_v(H_q) = 1$) at the conservation of both $G$ and $H_q$ graph's connectedness, it is necessary and sufficient for all the $H_q$ graph's vertexes to be simple.

Let's finally show that condition ($\delta_v(H_q) = 1$) appears to be both necessary and sufficient condition of the equality of both graph's cyclomatic numbers.

Let us assume that two connected $G$ and $H_q$ graphs have one initial binary $L = Q \times Q$ relation's system, all the $H_q$ graph's vertexes are simple, but $\nu(G) \neq \nu(H_q)$ at the same time. Let us consequently exclude from both graphs the same binary relations (the elements of the $L = Q \times Q$ set) at the condition of keeping the connectedness, reducing to zero one of the cyclomatic $\nu(G)$ or $\nu(H_q)$ numbers:

- If $\nu(G)$ is the first to reach zero, then it will prove for the $G$ graph to be a tree, and for the $H_q$ graph to have cycles.
- If $\nu(H_q)$ is the first to reach zero, then it will prove for the $H_q$ graph to be a tree, and for the $G$ graph to have cycles.

Both are impossible, because it contradicts the supposition that in the foundation of the graphs there is one initial binary $L = Q \times Q$ relation's system. Therefore, if the $G$ and $H_q$ graphs are connected, have one initial binary relation's system and $\delta_v(G) = \delta_v(H_q) = 1$, then: $\nu(G) = \nu(H_q)$

Well, if both graphs: $G$ – vertex, and $H_q$ – edge, have one and the same initial binary relation's $L = Q \times Q$ system, adjusted on the $Q = \{q_i\}$ set, then for the purpose of the cyclomatic numbers being equal, it is both necessary and sufficient for all the $H_q$ graph's vertexes be simple, in other words, for all the $r_{xy} = 1$ the condition ($\delta_v(G) = \delta_v(H_q) = 1$) must be satisfied.

The theorem is proved.

Let us agree to name the $\|r_{ij}\|_1^{n+s_q}$ matrix (an adjacency matrix of the $H_q$ graph's edges), which satisfy to condition (1) of the theorm 4, a canonical adjacency matrix and denote it as $R_k = \|r_{ij}\|_1^{n_k} = \|r_{ij}\|_1^{n+s_k}$. Let us also agree to name both $G$ and $H$ graphs, set by the canonical adjacency $R_k = \|r_{ij}\|_1^{n+s_k}$ matrix as the canonical both vertex and edge graphs and accordingly denote them as $G_k(Q_k, \Gamma_k)$ and $H_k(V_k, \mathrm{Q}_k)$. It is obvious that the $G_k(Q_k, \Gamma_k)$ graph is the adjacency graph for the $H_k(V_k, \mathrm{Q}_k)$ graph's edges, and the initial $G(Q, \Gamma)$ graph is the graph of the partial or the total transit adjacency for the $H_k(V_k, \mathrm{Q}_k)$ graph's edges.

Theorem 4 has evident corollaries.

### *Corollary 1*

If a canonical $G_k$ graph is the adjacency graph for the canonical $H_k$ graph's edges, then their cyclomatic numbers are equal: $\nu(G_k) = \nu(H_k)$.

### *Corollary 2*

If the $L$ matrix satisfies to the conditions of the theorem 3 *"On the equality of the cyclomatic numbers of both vertex and edge graphs"*, then it satisfies to the conditions of theorem 1.

### *Theorem 5 "An extension of the theorem "On the equality of the cyclomatic numbers of both vertex and edge graphs" for the case of the $p$-connected graphs"*

Let the $G$ graph, which has the $p$ components of the connectedness, be specified by the $E$ matrix – the vertex matrix, which corresponds to the quasicanonical $R_q$ matrix – the matrix of the $H_q$ graph's edges, which also has the $p$ components of the connectedness. For the purpose of equality of both graph's cyclomatic numbers, accordingly: $\nu_p(G_p) = \nu_p(H_p)$, it is both necessary and sufficient for all the $r_{xy} = 1$ elements to meet condition (1) of theorem 4.

From theorems 3, 4 and 5 it follows that:

### *Corollary 3*

If the canonical $G_k$ graph with the $p$ components of the connectedness is adjacency to the edge canonical $H_k$ graph also with the $p$ components of the connectedness, then their cyclomatic numbers are equal.

On the basis of the theorems 4 and 5 let us formulate, as evident, a theorem on the canonical binary relation's system.

### *Theorem 6 "On a canonical binary relation's system"*

Let the $Q = \{q_i\}$ set is given. It has the $p$ components of the connectedness, and the binary $L = Q \times Q$ relation's system, which has the cyclomatic $\nu_p(L)$ number.

A $Q = \{q_i\}$ set and a binary $L = Q \times Q$ relation's system, adjusted on this set, may be at the same time presented by the edge $H_k$ graph with the $p$ components of the connectedness and the cyclomatic $\nu_p(H_k) = \nu_p(L)$ number and its conjugate vertex $G_k$ graph with the $p$ components of the connectedness and the same cyclomatic $\nu_p(G_k) = \nu_p(L)$ number if and only if the binary $L = Q \times Q$ relation's system has the canonical form, in other words, the $\|l_{ij}\|_1^n = L$ matrix satisfies to theorem 4 conditions.

While proving the theorem 2, we were not interested in the concrete essence of the $q_x R q_y$ binary relation, but only the transitivity property was important, so the following theorem is evident.

## *Theorem 7 "On the binary relation's normalization"*

Any arbitrary $L = Q \times Q$ system of the binary $q_x R q_y$ relation's, which is assigned on the $Q = \{q_i\}$ set, may be brought to the canonical form by the way of the consistent application of the $\Delta n -$ transformation to that $\left\|l_{ij}\right\|_1^n = L = Q \times Q$ matrix's $l_{xy} = 1$ elements, which do not satisfy to theorem 4 conditions.

Let us give an example of bringing an arbitrary $L$ matrix to the canonical form, and the construction of the corresponding graphs – both vertex $G_k$ and edge $H_k$ graphs. An initial $L$ matrix is presented in fig. 26. The $l_{ij} = 1$ elements, which do not satisfy to theorem 4 conditions, are marked by the circles. On the initial $G$ graph the corresponding edges are marked by the criss-crosses.

$\left\|l_{ij}\right\|_1^8 =$

| | a | b | c | d | e | f | g | h |
|---|---|---|---|---|---|---|---|---|
| a | | 1 | (1) | 1 | | | | (1) |
| b | | | (1) | | (1) | | | (1) |
| c | | | | | | (1) | (1) | |
| d | | | | | | (1) | (1) | |
| e | | | | | | | | 1 |
| f | | | | | (1) | | | (1) |
| g | | | | | 1 | | | |
| h | | | | | | | | |

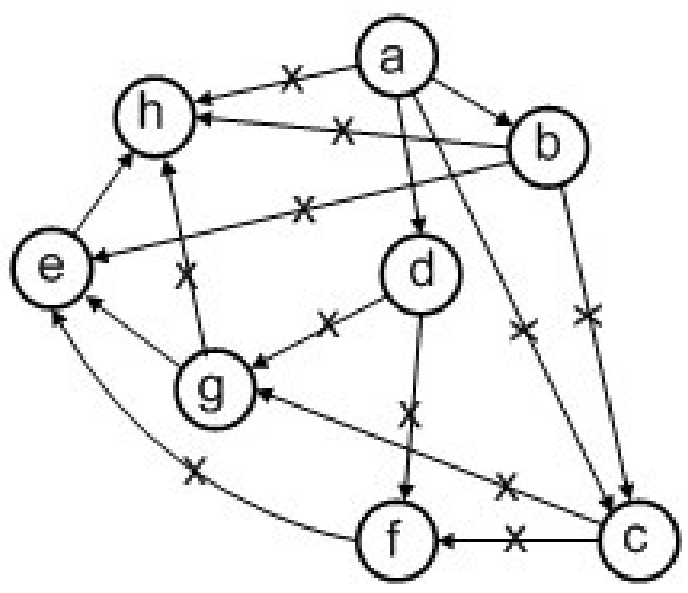

Fig. 26. An arbitrary $\boldsymbol{L}$ matrix and the $\boldsymbol{G}$ graph, corresponding to it.

After the application of the $\Delta n$-transformation to the marked elements, the matrix obtains the canonical $\left\|l_{ij}\right\|_1^{19} = \left\|r_{ij}^{(k)}\right\|_1^{19}$ form (fig. 27).

$\left\|r_{ij}^{(k)}\right\|_1^{19} =$

| | a | b | d | k | l | q | m | n | r | s | c | t | u | f | g | o | p | e | h | $N_{kn}$ | $N_{kk}$ |
|---|---|---|---|---|---|---|---|---|---|---|---|---|---|---|---|---|---|---|---|---|---|
| a | | 1 | 1 | 1 | 1 | | | | | | | | | | | | | | | 1 | 2 |
| b | | | | | | 1 | 1 | 1 | | | | | | | | | | | | 2 | 3 |
| d | | | | | | | | | 1 | 1 | | | | | | | | | | 2 | 4 |
| k | | | | | | | | | | | 1 | | | | | | | | | 2 | 5 |
| l | | | | | | | | | | | | | | | | | | | 1 | 2 | 11 |
| q | | | | | | | | | | | 1 | | | | | | | | | 3 | 5 |
| m | | | | | | | | | | | | | | | | | | 1 | | 3 | 10 |
| n | | | | | | | | | | | | | | | | | | | 1 | 3 | 11 |
| r | | | | | | | | | | | | | | 1 | | | | | | 4 | 7 |
| s | | | | | | | | | | | | | | | 1 | | | | | 4 | 8 |
| c | | | | | | | | | | | | 1 | 1 | | | | | | | 5 | 6 |
| t | | | | | | | | | | | | | | 1 | | | | | | 6 | 7 |
| u | | | | | | | | | | | | | | | 1 | | | | | 6 | 8 |
| f | | | | | | | | | | | | | | | | 1 | 1 | | | 7 | 9 |
| g | | | | | | | | | | | | | | | | | | 1 | | 8 | 10 |
| o | | | | | | | | | | | | | | | | | | 1 | | 9 | 10 |
| p | | | | | | | | | | | | | | | | | | | 1 | 9 | 11 |
| e | | | | | | | | | | | | | | | | | | | 1 | 10 | 11 |
| h | | | | | | | | | | | | | | | | | | | | 11 | 12 |

Fig. 27.

The two columns – $N_{hn}$ and $N_{hk}$ (the numbers of the initial and the final $H_k$ graph's $q_i$ vertexes) are attached on the right side of the $\left\|r_{ij}^{(k)}\right\|_1^{19}$ matrix.

A $\left\|r_{ij}^{(k)}\right\|_1^{19}$ matrix allows arranging the $\left\|f_{hg}\right\|_1^{12}$ matrix, which is adjacency to the $H_k$ graph's vertexes (fig. 28), and constructing the canonical graphs – both the $G_k$ vertex graph and the $H_k$ edge graph (fig. 29). The cyclomatic numbers of both graphs, presented in fig. 29a and 29b, are the same and are equal to 8.

$\left\|f_{hg}\right\|_1^{12} =$

| | 1 | 2 | 3 | 4 | 5 | 6 | 7 | 8 | 9 | 10 | 11 | 12 |
|---|---|---|---|---|---|---|---|---|---|---|---|---|
| 1 | | a | | | | | | | | | | |
| 2 | | | b | d | k | | | | | | l | |
| 3 | | | | | q | | | | | m | n | |
| 4 | | | | | | | r | s | | | | |
| 5 | | | | | | c | | | | | | |
| 6 | | | | | | | t | u | | | | |
| 7 | | | | | | | | | f | | | |
| 8 | | | | | | | | | | g | | |
| 9 | | | | | | | | | | o | p | |
| 10 | | | | | | | | | | | e | |
| 11 | | | | | | | | | | | | h |
| 12 | | | | | | | | | | | | |

Fig. 28. An adjacency $\left\|\boldsymbol{f_{hg}}\right\|_{\boldsymbol{1}}^{\boldsymbol{12}}$ matrix of the $\boldsymbol{H_k}$ graph's vertexes

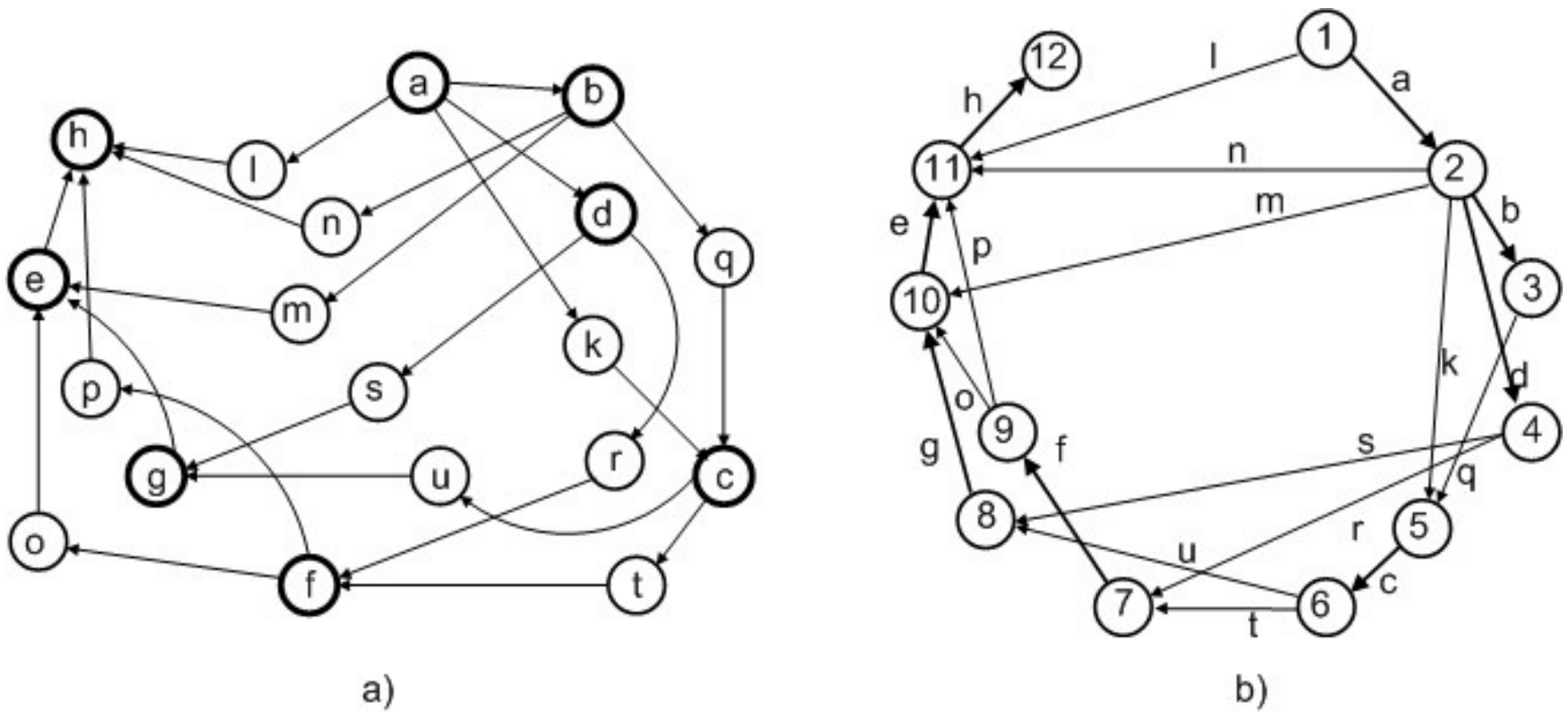

Fig. 29. The canonical graphs – the $\boldsymbol{G_k}$ vertex graph and the $\boldsymbol{H_k}$ edge graph

Theorems above allow formulating the following evident corollaries:

## *Corollary 4*

A vertex $G$ graph is the adjacency graph to the edges of some edge $H$ graph if and only if the adjacency matrix of the vertex $G$ graph has either the canonical or the quasicanonical form.

**Remark.** The consequence does not coincide with the mentioned above Krause condition [4] since Krause condition regards to the undirected graphs, and the corollary 4 – only to the directed graphs.

### *Corollary 5*

If in the $H(V, Q)$ graph, which has either the canonical or the quasicanonical form, we can pick out such Euler partial $H_e(V_e, Q_e)$ graph that for all of its vertexes $s_h(H_e) = 2$, and this graph also has all these edges, that have the one-to-one depentanizer to the initial $G(Q, \Gamma)$ graph's vertexes, then the $G(Q, \Gamma)$ graph has the Hamilton cycle, and this cycle can be determined unambiguously.

An example of Euler partial $H_e(V_e, Q_e)$ graph's correspondence and Hamilton cycles in the canonical graphs is presented in fig. 30 ÷ 35.

$\|l_{ij}\|_1^5 =$

| | A | B | C | D | E |
|---|---|---|---|---|---|
| A | | 1 | | 1 | 1 |
| B | 1 | | 1 | 1 | 1 |
| C | | 1 | | 1 | 1 |
| D | 1 | 1 | 1 | | |
| E | 1 | 1 | 1 | | |

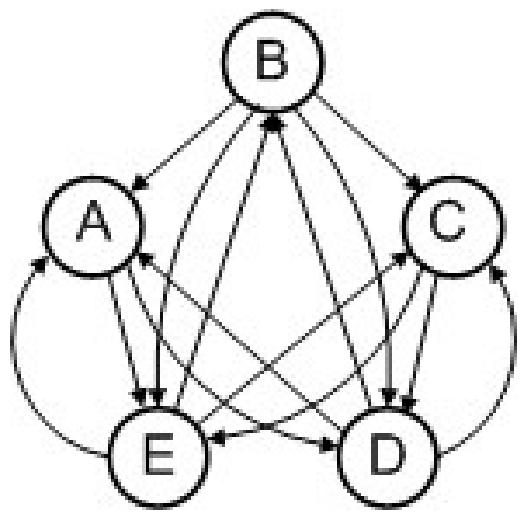

Fig. 30. Initial $\boldsymbol{L}$ matrix and the vertex $\boldsymbol{G(Q, \Gamma)}$ graph, corresponding to it.

$\|r_{ij}^{(k)}\|_1^{21} =$

| | A | H | F | G | E | W | V | Z | B | K | N | L | M | C | R | P | Q | D | S | T | U | $N_{hn}$ | $N_{hk}$ |
|---|---|---|---|---|---|---|---|---|---|---|---|---|---|---|---|---|---|---|---|---|---|---|---|
| A | | 1 | 1 | 1 | | | | | | | | | | | | | | | | | | 1 | 2 |
| H | | | | | 1 | | | | | | | | | | | | | | | | | 2 | 3 |
| F | | | | | | | | | 1 | | | | | | | | | | | | | 2 | 5 |
| G | | | | | | | | | | | | | | | | | | 1 | | | | 2 | 9 |
| E | | | | | | 1 | 1 | 1 | | | | | | | | | | | | | | 3 | 4 |
| W | 1 | | | | | | | | | | | | | | | | | | | | | 4 | 1 |
| V | | | | | | | | | 1 | | | | | | | | | | | | | 4 | 5 |
| Z | | | | | | | | | | | | | | 1 | | | | | | | | 4 | 7 |
| B | | | | | | | | | | 1 | 1 | 1 | 1 | | | | | | | | | 5 | 6 |
| K | 1 | | | | | | | | | | | | | | | | | | | | | 6 | 1 |
| N | | | | | 1 | | | | | | | | | | | | | | | | | 6 | 3 |
| L | | | | | | | | | | | | | | 1 | | | | | | | | 6 | 7 |
| M | | | | | | | | | | | | | | | | | | 1 | | | | 6 | 9 |
| C | | | | | | | | | | | | | | | 1 | 1 | 1 | | | | | 7 | 8 |
| R | | | | | 1 | | | | | | | | | | | | | | | | | 8 | 3 |
| P | | | | | | | | | 1 | | | | | | | | | | | | | 8 | 5 |
| Q | | | | | | | | | | | | | | | | | | 1 | | | | 8 | 9 |
| D | | | | | | | | | | | | | | | | | | | 1 | 1 | 1 | 9 | 10 |
| S | 1 | | | | | | | | | | | | | | | | | | | | | 10 | 1 |
| T | | | | | | | | | 1 | | | | | | | | | | | | | 10 | 5 |
| U | | | | | | | | | | | | | | 1 | | | | | | | | 10 | 7 |
| | 1 | 2 | 2 | 2 | 3 | 4 | 4 | 4 | 5 | 6 | 6 | 6 | 6 | 7 | 8 | 8 | 8 | 9 | 10 | 10 | 10 | | |

Fig. 31. A matrix after the $\boldsymbol{\Delta n}$–transformation

$\|f_{hg}\|_1^{10}$ -

| | 1 | 2 | 3 | 4 | 5 | 6 | 7 | 8 | 9 | 10 |
|---|---|---|---|---|---|---|---|---|---|---|
| 1 | | A | | | | | | | | |
| 2 | | | H | | F | | | | G | |
| 3 | | | | E | | | | | | |
| 4 | W | | | | V | | Z | | | |
| 5 | | | | | | B | | | | |
| 6 | K | | N | | | | L | | M | |
| 7 | | | | | | | | C | | |
| 8 | | | R | | P | | | | Q | |
| 9 | | | | | | | | | | D |
| 10 | S | | | | T | | U | | | |

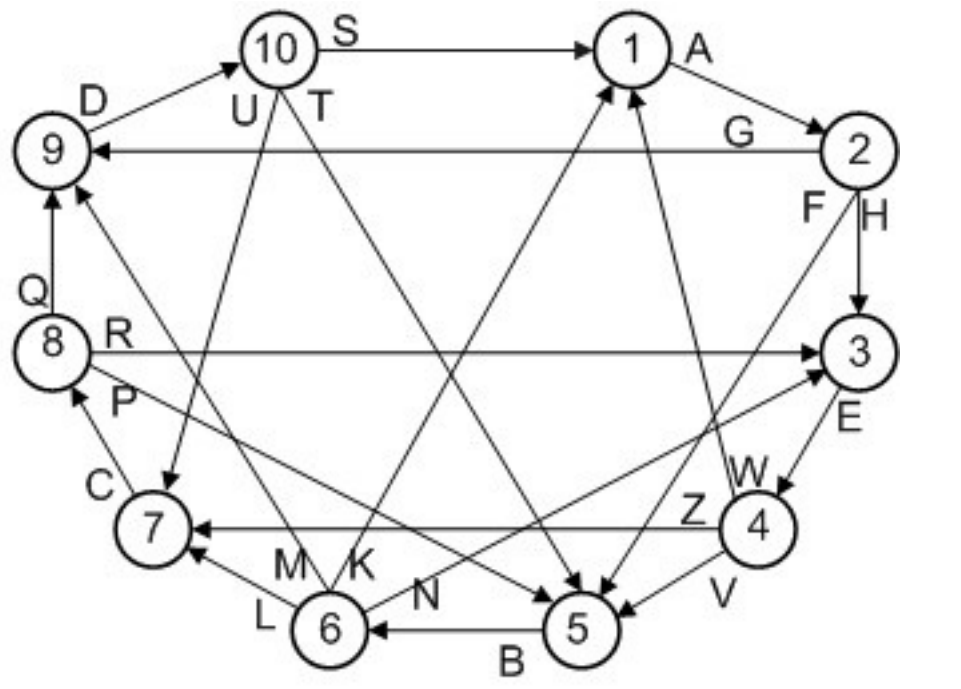

Fig. 32

Fig. 30÷32 does not need for the explanation.

The one of Euler partial $H_e(V_e, \mathrm{Q}_e)$ graphs and the Hamilton cycle AEBCDA corresponding to it are presented in fig. 33. Other examples are presented in fig. 34 and 35.

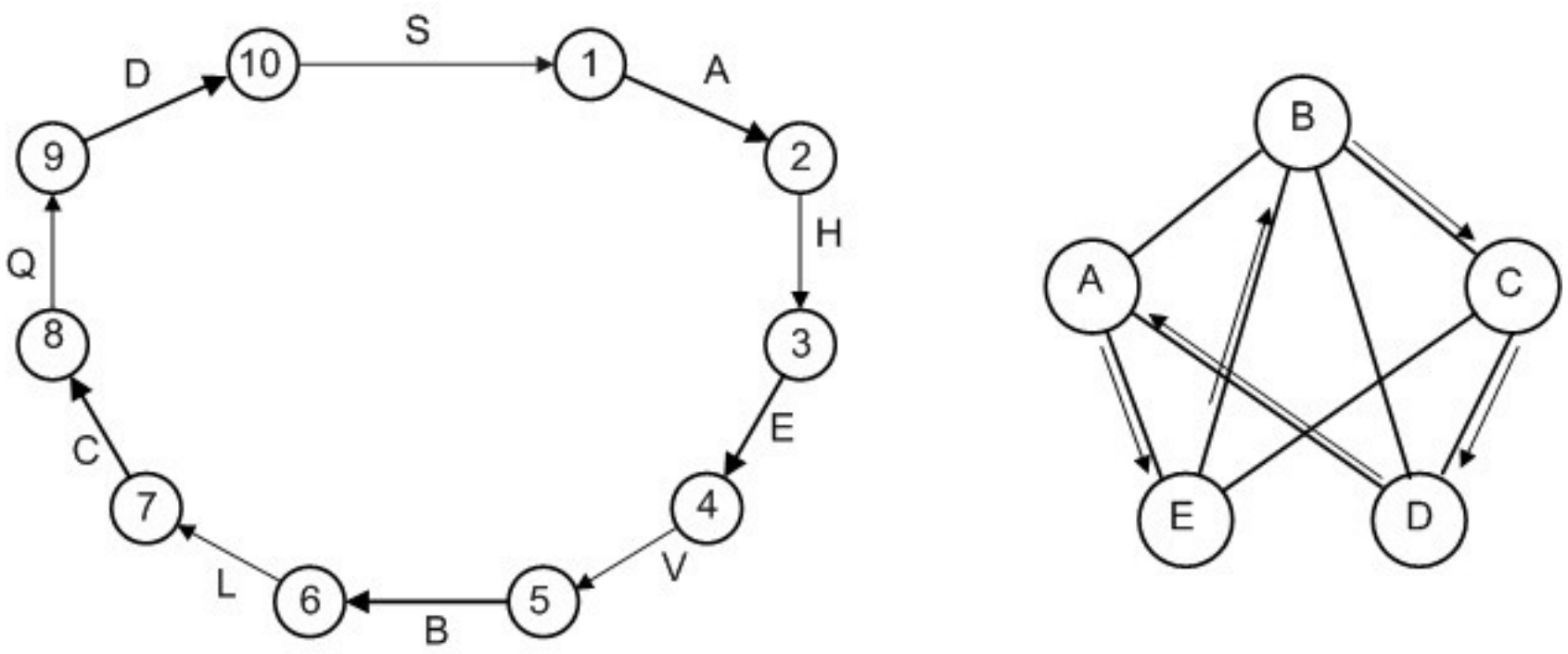

Fig. 33. One of Euler partial $\boldsymbol{H_e}(\boldsymbol{V_e}, \mathbf{Q_e})$ graphs and the Hamilton cycle AEBCDA corresponding to it.

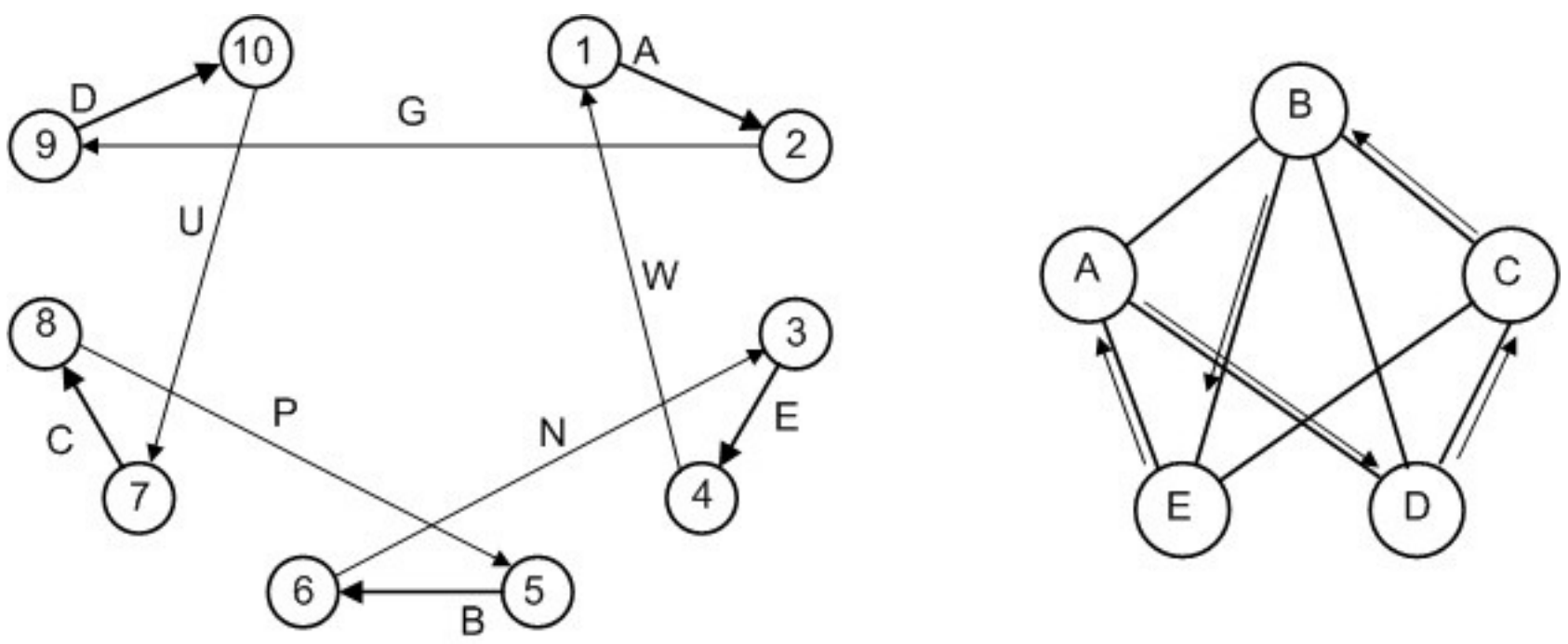

Fig. 34.

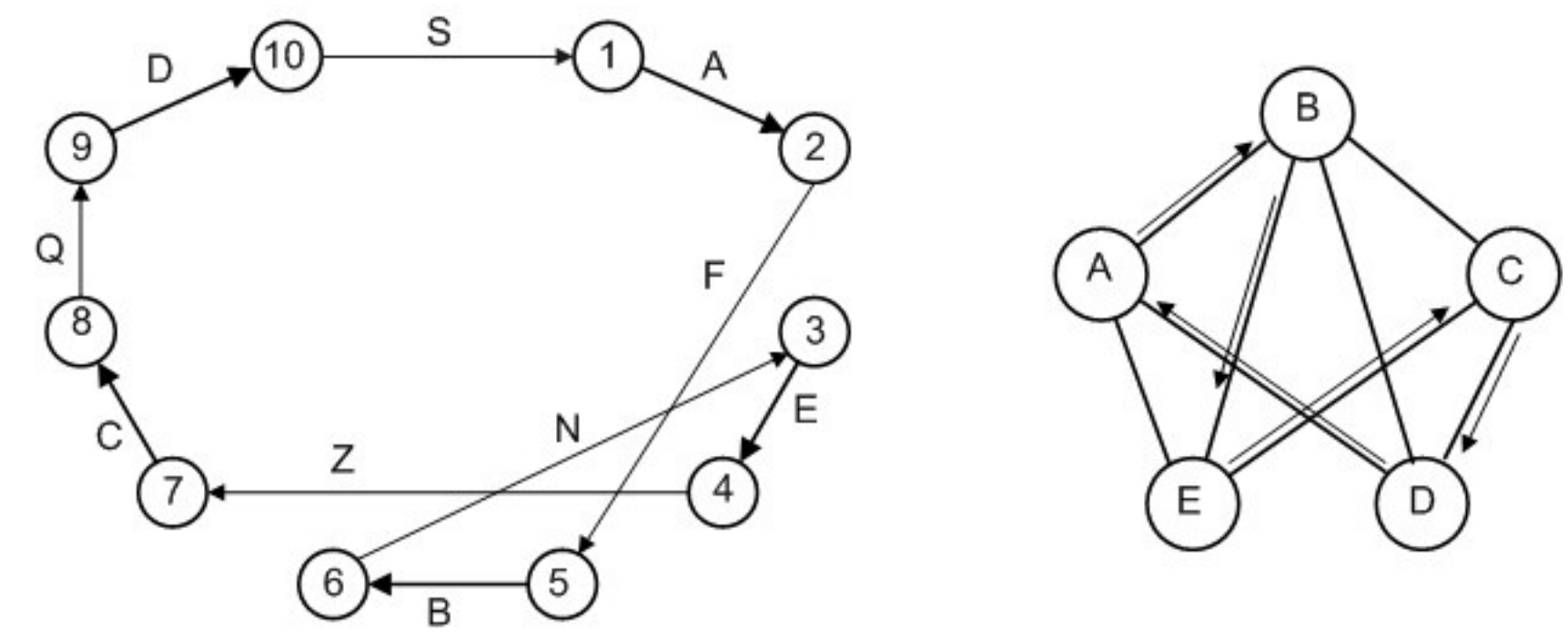

Fig. 35.

### *Corollary 6*

A number of Hamilton cycles in the $G$ graph is equal to the number of Euler partial graphs in either the $H_q$ graph or the $H_k$ graph, obligatory containing also all the edges that are the one-to-one depentanizer to the $G$ graph's vertexes.

### *Corollary 7*

If either the $H_q$ graph or the $H_k$ graph has a family of the noncrossing by the edges the circuits (paths) that cover all those edges, with which the $G$ graph's vertexes have the one-to-one depentanizer, then to this family of the circuits (paths) in the $G$ graph corresponds a family of the noncrossing similar either circuits or chains, which cover all the $G(Q,\Gamma)$ graph's vertexes.

## 5 Forming vertex graphs

The operation of the adjacency matrix's normalization, which is based on the $\Delta n$–transformation of the binary relation, increases the initial $L$ matrix's degree and transforms the initial $Q$ set into the main set of either the canonical or the quasicanonical graph.

Let's examine a contrary operation, which will permit either to decrease a degree of the initial matrix or to reduce the matrix without breaking a binary relation's system between the elements of some set, which we'll denote as the forming set.

**Definitions:**

**5.1. A $(-\mathbf{\Delta n})$–transformation.** By the $(-\Delta n)$-transformation of the $\left\|l_{ij}\right\|_1^n$ matrix we'll agree to understand an exclusion of one $q_\alpha$ row (both the line and the column) from the matrix with the simultaneous replacement of both $l_{x\alpha}=1$ and $l_{\alpha y}=1$ elements, if such a pair exists, by one $l_{xy}=1$ element at the condition that the replaceable elements satisfy to the conditions of theorem 4 (the theorem on the canonical adjacency matrix). We mean condition (1).

An adjacency $\left\|r_{ij}^{(k)}\right\|_1^{10}$ matrix, which satisfies to theorem 4 conditions, is presented in fig. 36a. A total number of the $r_{ij} = 1$ elements is equal 13. The 12 of them get into pairs; each pair, taken separately, may be replaced by one element. For example: the pair $r_{ac} = 1$ and $r_{cg} = 1$, corresponding to the $a < c < g$ relation may be replaced by one $l_{ag} = 1$ element. It gives us a $a < g$ relation, which is equal to the $a < c < g$ relation for both $a$ and $g$ elements. At the same time the $c$ row in the $\left\|r_{ij}^{(k)}\right\|_1^{10}$ matrix is excluded. A pair $r_{bf} = 1$ and $r_{fa} = 1$ also may be replaced by one $l_{ba} = 1$ element with the simultaneous excluding of the $f$ row from the $\left\|r_{ij}^{(k)}\right\|_1^{10}$ matrix, and so forth. An $r_{ab} = 1$ element does not enter the pairs of such kind.

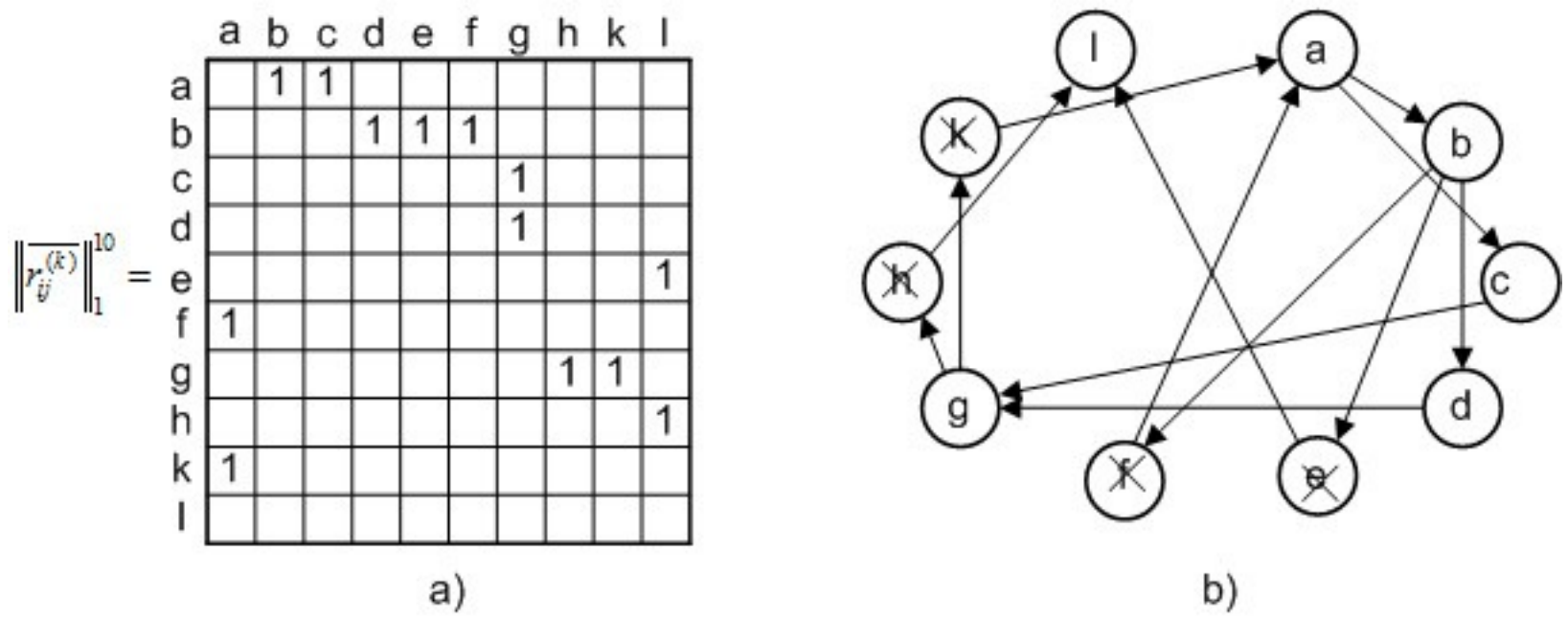

Fig. 36. The adjacency $\left\|\overline{\boldsymbol{r}_{ij}^{(k)}}\right\|_1^{10}$ matrix, which satisfies to the conditions of theorem 4.

**5.2. A forming binary relation's system.** Let us settle to denote as a forming one a system $L^* = Q^* \times Q^*$ of binary $q_i R q_j$ relations, which is adjusted on the $Q^* \subset Q$ set or adjusted, concerning the elements $q_i^* \in Q^* \subset Q$, on such relation's system, which in the $L^* = \left\|l_{ij}^*\right\|_1^n$ matrix has not a single pair of elements ($l_{x\alpha} = 1$ and $l_{\alpha y} = 1$), which satisfies to the conditions of theorem 4.

**5.3. A forming set.** If a $L^* = Q^* \times Q^*$ system of binary $q_i R q_j$ relations, adjusted on the $Q^*$ set, is the forming one, then let us settle to denote the $Q^* = \{q_i^*\}$ set as a forming set.

**5.4. A forming vertex graph.** By the forming vertex graph we'll denote the directed $G^*(Q^*, \Gamma^*)$ graph, which has no such a $q$ vertex, for which $\sum \rho(q_i) = 0$ and $\sum |\rho(q_i)| = 2$, that means the absence of elementary vertexes.

A binary $q_i R q_j$ relation's $L = Q \times Q$ system, arbitrary adjusted on the connected $Q = \{q_i\}$ set, allows constructing the connected $G(Q, \Gamma)$ graph, which vertexes are $q_i \in Q$ and have $\sum |\rho(q_i)| \geq 1$.

Either the normalization or the quasinormalization of the binary relation consists of including into the $G(Q, \Gamma)$ graph the additional vertexes, which have $\sum |\rho(q_i)| = 2$, and the additional edges. Since the $\Delta n$-transformation is the conservative one, then it is the reversible one. So we may apply the $(-\Delta n)$- transformation to every pair $q_x R q_\alpha$ and $q_\alpha R q_y$ of the relation, replacing it by the $q_x R q_y$ relation.

Let both $Q$ and $Q^* \subset Q$ sets be given. Also let the binary relation's $L = Q \times Q$ system be given. It allows constructing the $G(Q, \Gamma)$ graph.

Let further for all the $q_i^* \in Q^*$, $\sum|q_i^*| \neq 2$, and for all the $q_i \in (\overline{Q \cap Q^*})$: $\sum|\rho(q_i)| = 2$ and $\sum \rho(q_i) = 0$. It is evident that we can exclude all the $q_i \in (\overline{Q \cap Q^*})$, which have $\sum|\rho(q_i)| = 2$, from the $G(Q, \Gamma)$ graph by applying the $(-\Delta n)$-transformation to them. Under such circumstances the $Q$ set will be reduced to the $Q^* = \{q_i^*\}$ set. A received binary relation's $L^* = Q^* \times Q^*$ system will be equivalent to the $L = Q \times Q$ system, concerning the elements of the $Q^* \subset Q$ set, owing to the transitivity property of the binary relation. A $G(Q, \Gamma)$ graph will be transformed into the $G^*(Q^*, \Gamma^*)$ graph. In such a way, we have a theorem of the reduction on the binary relation's system.

## *Theorem 8: "A theorem on the reduction of the binary relation's system"*

If the arbitrary, given on the connected $Q = \{q_i\}$ set, binary $q_i R q_j$ relation's $L = Q \times Q$ system contains such pairs of the relations as $q_x R q_\alpha$ and $q_\alpha R q_y$, which in the $\|l_{ij}\|_1^n$ matrix correspond to the pairs $l_{x\alpha} = 1$ and $l_{\alpha y} = 1$ of the elements, which satisfy to the conditions of theorem 4, then such $L = Q \times Q$ system does not appear to be the forming system, but always may be, by applying the $(-\Delta n)$-transformation to the mentioned pairs, reduced to the forming $L^* = Q^* \times Q^*$ system with reducing of the $Q$ set to the forming $Q^*$ set at the same time. It may be also shown that theorem 8 is also right for the $p$-connected set.

### *Corollary 8*

A heterogeneous directed $G(Q, \Gamma)$ graph, which $q_i$ vertexes may have $\sum|\rho(q_i)| \geq 1$, always may be transformed into the forming $G^*(Q^*, \Gamma^*)$ graph by the way of the replacement in the $G(Q, \Gamma)$ graph every $q_\alpha$ vertex, which has $\sum \rho(q_\alpha) = 0$ and $\sum \rho(q_\alpha) = 2$, and both $(q_x, q_\alpha)$ and $(q_\alpha, q_y)$ edges by one $(q_x, q_y)$ edge. At the same time new binary relations system between the residuary $q_i^* \in Q^*$ vertexes will be equivalent to the initial relation's system among such vertexes according to transitivity property.

### *Corollary 9*

A forming $G^*(Q^*, \Gamma^*)$ graph's cyclomatic $\nu(Q^*)$ number is equal to the initial $G(Q, \Gamma)$ graph's cyclomatic number.

The acceptance of such terms as a forming graph and a forming system of the binary relations becomes evident from the definition of the operation of the reduction (the $(-\Delta n)$-transformation) and theorem 8. All the elementary vertexes are excluded from the $G$ graph at the reduction of the $L$ matrix. A normalization of the previously reduced $L^*$ matrix allows achieving either $H_k$ or $H_q$ graphs without the elementary vertexes. Similar operation may be applied to the $F$ matrix. In this case the elementary vertexes will be excluded from either $H_k$ or $H_q$ graphs.

If, for practical purposes, we need a construction of the ordinary net model with the stochastic structure, then in such model the elementary vertexes do not contain the logical operations, so they are not needed for the forming of the model's logical structure. An operation of reduction allows reducing the initial set for such marginally number of the elements that still guarantees the forming of the model's logical structure, being adequate to the logical structure of the process. That's why the operation of reduction undoubtedly allows constructing the models of the optimal structure.

Once again let us examine the transformation of either the $H_k$ graph or the $G_k$ graph into the forming $G^*$ graph. Fig. 37b represents the $H_k$ graph and the adjacency $F_k$ matrix of its vertexes (fig 37a). The $H_k$ graph corresponds to the adjacency $R_k$ matrix (fig 36a) of the $H_k$ graph's edges and also to the vertex $G_k$ graph (fig. 36b). From all the $G_k$ graph's vertexes the 6 ($q_c$; $q_a$; $q_e$; $q_f$; $q_h$; $q_k$) vertexes have $\sum \rho(q_i) = 0$ and $\sum|\rho(q_i)| = 2$. So, the $G_k$ graph may be reduced and transformed into the forming graph.

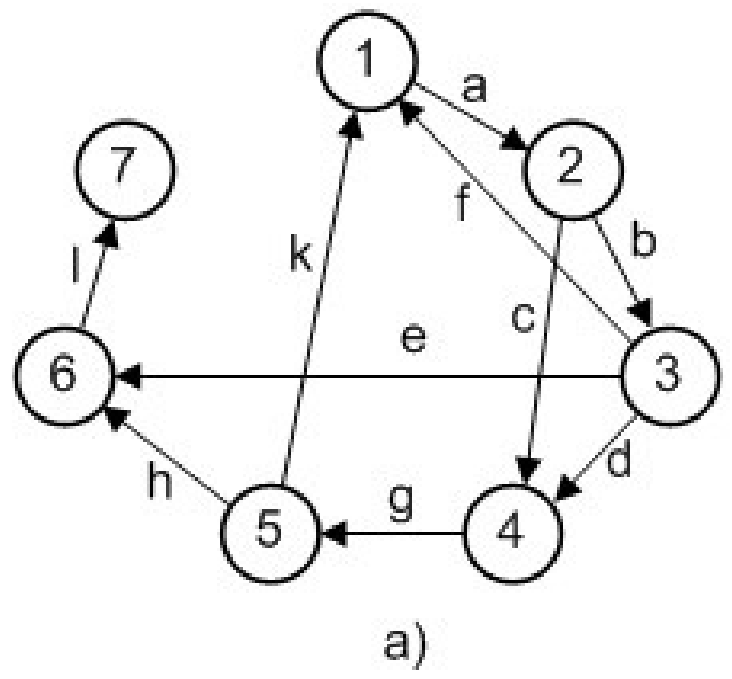

$\|f_{hq}\|_1^7 =$

| | 1 | 2 | 3 | 4 | 5 | 6 | 7 |
|---|---|---|---|---|---|---|---|
| 1 | \ | a | | | | | |
| 2 | | \ | b | c | | | |
| 3 | f | | \ | d | | e | |
| 4 | | | | \ | g | | |
| 5 | k | | | | \ | h | |
| 6 | | | | | | \ | l |
| 7 | | | | | | | \ |

b)

Fig. 37. An edge $\boldsymbol{H_k}$ graph and the adjacency $\boldsymbol{F_k}$ matrix of its vertexes

The matrix reducing is represented in fig. 38 and consists in the following:

1. The values $\sum_{i/j} l_{ij}$ and $\sum_{j/i} l_{ij}$ are calculated.
2. The values of $\sigma_{ij} = \left(\sum_{\frac{i}{j}} l_{ij}\right) * (\sum_{\frac{j}{i}} l_{ij})$ are indicated for all the $i = j$ along the main matrix's diagonal (this values are circled in fig. 39a).
3. If $\sigma_{ij} = \left(\sum_{\frac{i}{j}} l_{ij}\right) * \left(\sum_{\frac{j}{i}} l_{ij}\right) = 1$, then the corresponding row is excluded from the matrix. At the same time both $l_{x\alpha} = 1$ and $l_{\alpha y} = 1$ elements are replaced by the $l_{xy} = 1$ element.

The excluded from the matrix elements are obliterated, and new included ones are bordered with the small squares. The rows, which are beyond any exclusion, are marked by the starlets. As a result we have the forming matrix (fig. 38b) and the forming graph (fig. 38c).

A forming set contains only four elements instead of ten in the initial set.

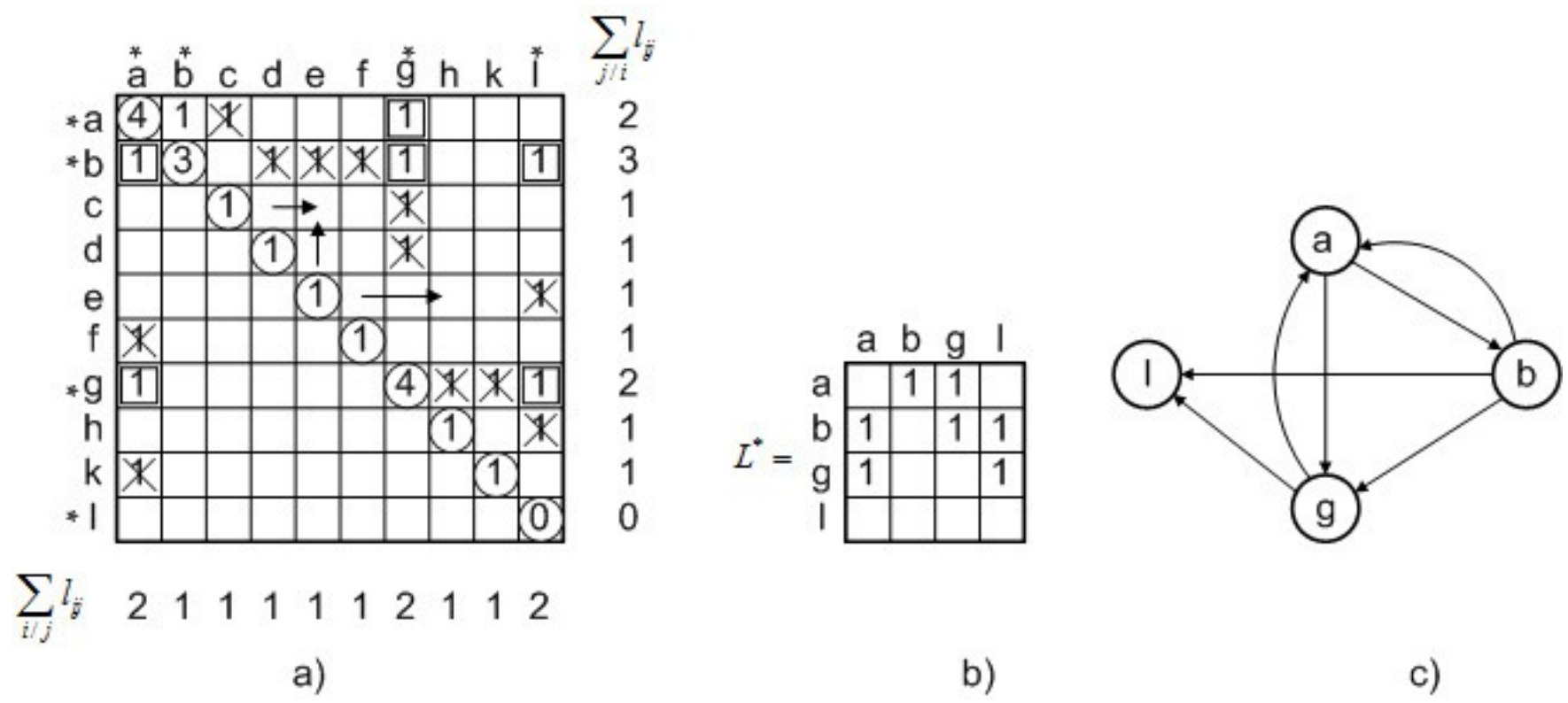

Fig. 38.

The cyclomatic numbers of all three graphs are the same and are equal to 4. It is easy to make sure that the binary relation's system on the forming set in all three graphs is the same.

In this example:

Table 1

| The row excluded | The pair replaced | By the element |
|---|---|---|
| $c$ | $r_{ac}=1$ and $r_{cg}=1$ | $l_{ag}=1$ |
| $d$ | $r_{bd}=1$ and $r_{dg}=1$ | $l_{bg}=1$ |
| $e$ | $r_{be}=1$ and $r_{el}=1$ | $l_{bl}=1$ |
| $f$ | $r_{ef}=1$ and $r_{fa}=1$ | $l_{ea}=1$ |
| $h$ | $r_{gh}=1$ and $r_{hl}=1$ | $l_{gl}=1$ |
| $k$ | $r_{gk}=1$ and $r_{ka}=1$ | $l_{ga}=1$ |

# 6 Conclusions

As a result of the research of the direct path matrix's (binary relation's matrix) properties both necessary and sufficient conditions of both the existence and the uniqueness of such direct path matrix's duality are proved. This matrix, at the same time, is both the adjacency vertex matrix of the directed vertex graph and the adjacency edge matrix of directed edge graph for the cases, when the cyclomatic numbers of both the vertex and the edge graphs are different (the quasicanonical adjacency matrix) and when the graph's cyclomatic numbers are equal (the canonical adjacency matrix). At that the vertex graph always is the adjacency edge's graph of the edge graph.

The proved theorems allow solving a problem of the transformation of the arbitrary (both directed and undirected) vertex graphs into directed edge graphs.

The proved theorems allow formulating the following foundations of the graph's isomorphism.

- The principle of the confluent duality
- The principle of the duality
- The principle of the normalization
- The principle of the reduction

## A principle of the confluent duality

The principle of the confluent duality (the quasiduality) is formulated on the base of the theorem on the quasicanonical adjacency matrix and determine both necessary and sufficient conditions of the duality of both directed and undirected both vertex and edge graphs, does not obligatory having the equal cyclomatic numbers.

So as the elements of the connected $Q = \{q_i\}$ set, on which the binary relation's $L = Q \times Q$ system in the $q_i R q_j$ form is adjusted, were at the same time both the edges of the connected $H_q(V_q, \mathrm{Q}_q)$ graph and the vertexes of the connected $G_q(Q_q, \Gamma_\mathrm{q})$ graph (the adjacency graph of the $H_q(V_q, \mathrm{Q}_q)$ graph's edges), it is both necessary and sufficient for the binary relation's $L = Q \times Q$ system to have the quasicanonical (the quasinormal) form, in other words, the $L$ matrix must satisfy to the conditions of the theorem on the quasicanonical adjacency matrix.

So, at the graphs transforming, the structural similarity between them becomes formed by the following relations between the corresponding similarity criteria:

- The numbers of the elements of the initial both $Q_0$ and $Q$ sets are equal, that is: $n_{Q_0} = n_Q$;
- The binary relation's systems on both $Q_0$ and $Q$ sets are equal, that is $L_0 = L$;
- The cyclomatic numbers of both the initial vertex $G$ graph and the edge $H_q$ graph may be different, that are: $\nu(G_q) \geq \nu(H_q)$.

## The principle of the strict duality

The duality principle is formulated on the base of the theorem on the canonical adjacency matrix and determines both the necessary and the sufficient conditions of the duality of both the vertex and the edge graphs, having the equal cyclomatic numbers.

The elements of the connected $Q = \{q_i\}$ set, on which the binary $q_i R q_j$ relation's $L = Q \times Q$ system (a cyclomatic $\nu_p(L)$ number) is assigned, are, at the same time, both the edges of the connected $H_k(V_k, \mathrm{Q}_k)$ graph and the vertexes of the connected $G_k(Q_k, \Gamma_\mathrm{k})$ graph (an adjacency graph of the $H_k(V_k, \mathrm{Q}_k)$ graph's edges), having the cyclomatic numbers accordingly $\nu_p(H_k)$ and $\nu_p(G_k)$, the same and equal $\nu_p(L)$, if and only if the binary relation's $L = Q \times Q$ system has the canonical (the normal) form, that is the $L$ matrix satisfies to the conditions of theorem 4.

So, the structural similarity between the graphs at their transformation, clarify itself by the equality of all three similarity criteria, that is:

- $n_{Q_0} = n_Q$;
- $L_0 = L$;
- $\nu(G) = \nu(H_k)$.

## The principle of the normalization

The principle of the normalization is formulated on the base of the theorems on both the quasinormalization and the normalization of the arbitrary adjacency matrixes and determines the single method of the transformation of every arbitrary matrix of the direct paths (the binary relation's matrix) to either the quasicanonical or the canonical form.

Every arbitrary $L \subset Q \times Q$ system of binary $q_i R q_j$ relation's form, adjusted on the $Q = \{q_i\}$ set, may be brought to either the quasicanonical (the quasinormal) or the canonical (the normal) form with the help of the single structural method of the $\Delta n$–transformation (the operation of either quasinormalization or normalization).

***Comments:***

1. In the general case only that $l_{ij} = 1$ elements of the $L \subset Q \times Q$ matrix, which do not satisfy to the conditions of theorem 4, are subjected to the $\Delta n$–transformation (the normalization).
2. If necessary, all the $l_{ij} = 1$ elements of the $L$ matrix may be subjected to the normalization (the reductive normalization). It is obvious that the binary relation's system between the elements of the initial set would not be broken at that.

## The reduction principle

The reduction principle is formulated on the base of the theorem on the forming both binary relation's system and set, and determines a single method of the transformation of the canonical, the quasicanonical or the arbitrary adjacency matrixes to the appearance of the forming adjacency matrixes.

If the $L$ matrix, corresponding to the binary relation's $L \subset Q \times Q$ system, adjusted on the $Q = \{q_i\}$ set, contains such pairs of the $l_{x\alpha} = 1$ and $l_{\alpha y} = 1$ elements, which satisfy to the conditions of theorem 4, then the $L \subset Q \times Q$ system is not obligatory the forming, but always may be converted to the appearance of the forming binary relation's $L^* \subset Q^* \times Q^*$ system with the help of the $(-\Delta n)$–transformation (the operation of the reduction).

The reduction principle may be applied for transforming the canonical adjacency matrixes to the quasicanonical form.

The reduction operation coupled with either the operation of the normalization (the quasinormalization) or the operation of the reductive normalization allows constructing either the vertex graphs with the optimal structure or the quasioptimal graphs, in other words, such graphs that does not contain the elementary vertexes.

When reforming the graphs into the graphs of the optimal structure the structural similarity of the equivalent graphs clarify itself by the equality of such similarity criteria:

- The numbers of the elements of the forming sets, that is: $n^*_{Q_0} = n^*_Q$;
- The binary relation's systems, adjusted on the forming sets, that is: $L^*_0 = L^*$;
- The cyclomatic numbers of either the initial vertex graph, the forming vertex graph or, constructed on its base, the canonical edge graph, that is: $\nu(G) = \nu(G^*) = \nu(H^*_k)$.

At constructing the edge graphs of the quasioptimal structure the cyclomatic numbers are connected with the correlation: $\nu(G) = \nu(G^*) \geq \nu(H^*_k)$.

At transforming the vertex graphs into the edge graphs such an operation as the construction of the $F$ matrix with the help of either canonical or quasicanonical $R$ matrix is used. This operation is undoubtedly interesting when constructing the net models of such systems, which complexity may be bind with the cyclomatic number.

# 7 Aknowledgments

Many thanks for help to improve this article to my friend Olga Volgina, who for several years had reformed my English. Special thanks for the members of my family who undergone all the difficulties side by side with me and who encouraged me in my work. Thanks to Michael Trofimov for his help in performing the material to arXiv.

Also I will be very grateful to those readers, who will find and send me a word about the uncovered misprints in order to improve the text. The matter is that on the base of the proved theorems were made the polynomial algorithms. It was also proved that the algorithms are local. And, finally, these algorithms were brought to the view of the programs.

# 8 Some designations

$G(Q, \Gamma)$ or $Q$ – the directed vertex graph
$H(Q, V)$ or $H$ – the directed edge graph
$Q$ – the set of vertexes
$\vec{Q}$ – the set of edges
$Q$ – the initial set of vertexes of the vertex graph or the initial set of edges of the edge graph
$\Gamma$ – set of edges of vertex graph
$V$ – set of vertexes of edge graph
the fundamental set
$L \subset Q \times Q$ – the binary relation's system

$L = \|l_{ij}\|_1^n$ – an arbitrary adjacency matrix of binary relations
$l_{ij}$ – an element of the arbitrary adjacency matrix of binary relations
$q_i$ – a vertex of the $G$ graph or an edge of the $H$ graph
$E = \|e_{ij}\|_1^n$ – adjacency matrix of the $G$ graph's vertexes
$R = \|r_{ij}\|_1^n$ – adjacency matrix of the $H$ graph's edges
$F = \|f_{hg}\|_1^m$ – the adjacency matrix of $H$ graph's vertexes
$G_q(Q_q, \Gamma_q)$ – the quasicanonical vertex graph
$H_q(Q_q, V_q)$ – the quasicanonical edge graph
$\|l_{ij}\|_1^{n_q}$ – the quasicanonical adjacency matrix
$\|e_{ij}\|_1^{n_q}$ – the $\mathrm{G_q}(\mathrm{Q_q}, \Gamma_\mathrm{q})$ graph's vertex adjacency matrix
$v_h$ – a vertex of the edge graph
$\nu(G_q)$ – the $\mathrm{G_q}$ graph's cyclomatic number
$\nu(H_q)$ – the $\mathrm{H_q}$ graph's cyclomatic number
$\|e_{ij}\|_1^{n_q}$ – quasicanonical adjacency matrix of vertexes of the $G_q(Q_q, \Gamma_q)$ graph.
$\|r_{ij}\|_1^{n_q}$ – quasicanonical adjacency matrix of edges of the $H_q(V_q, Q_q)$ graph.
$\nu$ – cyclomatic number
$R(q_i; q_j)$ – the binary relation's system
$n$ – degree of matrix or a number of elements of the initial set
$F_q$ – quasicanonical adjacency matrix of vertexes of the $H_q$ graph or operator adjacency matrix
$R_q$ – quasicanonical adjacency matrix of vertexes of the $H_q$ graph
$k$ – a number of connections which enter the vertex
$p$ – a number of connections which come out of the vertex
$G_k = G_k(Q_k, \Gamma_k)$ – canonical vertex graph
$H_k = H_k(V_k, Q_k)$ – canonical edge graph
$p$ – a property of the graph's connectivity
$q_x R q_y$ – binary relation
$H_э = H_э(V_э, Q_э)$ – Euler partial graph
$L^* = Q^* \times Q^*$ – A forming binary relation system
$Q^* = \{q_i^*\}$ – A forming set
$G^*(Q^*, \Gamma^*)$ – A forming vertex graph
$N_{beg}$ – номера начальных событий
$N_{fin}$ – номера конечных событий